\documentclass[aoas,preprint]{imsart}

\RequirePackage[OT1]{fontenc}
\RequirePackage{amsthm,amsmath}
\RequirePackage{natbib}
\RequirePackage[colorlinks,citecolor=blue,urlcolor=blue]{hyperref}

\usepackage{graphicx}

\usepackage{subfigure} 
\usepackage{array} 

\usepackage{bm}
\usepackage{amssymb}

\usepackage{booktabs}
\usepackage{url}
\usepackage{geometry}

\usepackage{verbatim}

\usepackage[usenames,dvipsnames]{xcolor}

\definecolor{dark-red}{rgb}{0.4,0.15,0.15}
\definecolor{dark-blue}{rgb}{0.15,0.15,0.4}
\definecolor{medium-blue}{rgb}{0,0,0.5}
\hypersetup{
    colorlinks, linkcolor={dark-blue},
    citecolor={dark-blue}, urlcolor={medium-blue}
}

\arxiv{1402.3580}

\startlocaldefs
\numberwithin{equation}{section}
\theoremstyle{plain}

\endlocaldefs

\begin{document}

\begin{frontmatter}
\title{Bayesian Inference for NMR Spectroscopy with Applications to Chemical Quantification}
\runtitle{Bayesian Inference for NMR Spectroscopy}

\begin{aug}
\author{\fnms{Andrew Gordon} \snm{Wilson}\thanksref{m1}\ead[label=e1]{agw38@cam.ac.uk}},
\author{\fnms{Yuting} \snm{Wu}\thanksref{m1}\ead[label=e2]{yw316@cam.ac.uk}},
\author{\fnms{Daniel J.} \snm{Holland}\thanksref{m1}\ead[label=e3]{djh79@cam.ac.uk}},
\\
\author{\fnms{Sebastian} \snm{Nowozin}\thanksref{m2}\ead[label=e4]{Sebastian.Nowozin@microsoft.com}},
\author{\fnms{Mick D.} \snm{Mantle}\thanksref{m1}\ead[label=e5]{mdm20@cam.ac.uk}},
\author{\fnms{Lynn F.} \snm{Gladden}\thanksref{m1}\ead[label=e6]{lfg1@cam.ac.uk}},
\author{\fnms{Andrew} \snm{Blake}\thanksref{m2}\ead[label=e7]{Andrew.Blake@microsoft.com}}

\runauthor{Wilson et al.}

\affiliation{Cambridge University\thanksmark{m1} and Microsoft Research\thanksmark{m2}}

\address{%
Andrew Gordon Wilson\\
Dept. of Engineering\\
University of Cambridge\\
Cambridge, United Kingdom\\
\printead{e1}}

\address{%
Yuting Wu\\
Dept. of Chemical Engineering and Biotechnology\\
University of Cambridge\\
Cambridge, United Kingdom\\
\printead{e2}}

\address{%
Daniel J. Holland\\
Dept. of Chemical Engineering and Biotechnology\\
University of Cambridge\\
Cambridge, United Kingdom\\
\printead{e3}}

\address{%
Sebastian Nowozin\\
Microsoft Research\\
21 Station Road\\
CB12FB Cambridge, United Kingdom\\
\printead{e4}}

\address{%
Mick D. Mantle\\
Dept. of Chemical Engineering and Biotechnology\\
University of Cambridge\\
Cambridge, United Kingdom\\
\printead{e5}}

\address{%
Lynn F. Gladden\\
Dept. of Chemical Engineering and Biotechnology\\
University of Cambridge\\
Cambridge, United Kingdom\\
\printead{e6}}

\address{%
Andrew Blake\\
Microsoft Research\\
21 Station Road\\
CB12FB Cambridge, United Kingdom\\
\printead{e7}}
\end{aug}

\begin{abstract}
Nuclear magnetic resonance (NMR) spectroscopy exploits the magnetic properties of atomic nuclei to
discover the structure, reaction state and chemical environment of molecules. 
We propose a probabilistic generative model and inference procedures for 
NMR spectroscopy.  Specifically, we use a weighted sum of trigonometric
functions undergoing exponential decay to model free induction decay (FID) signals.  We discuss the challenges 
in estimating the components of this general model -- amplitudes, phase shifts, frequencies, decay rates, and
noise variances -- 
and offer practical solutions.  We compare with conventional Fourier transform spectroscopy for estimating 
the relative concentrations of chemicals in a mixture, using synthetic and experimentally acquired FID signals.
We find the proposed model is particularly robust to low signal to noise ratios (SNR), and overlapping peaks in the 
Fourier transform of the FID, enabling accurate predictions (e.g., $1$\% sensitivity at low SNR) 
which are not possible with conventional spectroscopy ($5$\% sensitivity).  
\end{abstract}



\end{frontmatter}


\section{Introduction}
\label{sec: intro}

Nuclear magnetic resonance (NMR) spectroscopy has greatly advanced our understanding of molecular properties, and is now 
widespread in analytical chemistry.  The theory of nuclear magnetic resonance postulates 
that protons and neutrons behave like gyroscopes that spin about their axes, generating their own small magnetic fields.  These concepts 
were first described by \citet{rabi1939molecular}, for which Isidor Rabi was awarded the 1944 Nobel prize in physics.  Later, \citet{bloch1946}
and \citet{purcell1946} showed how NMR could be used to understand the structure of molecules in liquids and solids, for which they shared the 
Nobel prize in physics in 1952.  Richard Ernst then won the 1991 Nobel prize in chemistry for developing Fourier transform NMR spectroscopy\footnote{See \citet{ernst1992nuclear} for a comprehensive review.}, which led to the prevalence of NMR as an analytic technique.  NMR spectroscopy
is well suited to studying both organic and inorganic molecules, including proteins, and other biochemical species \citep{Barrett2013}, and 
is routinely used to identify the structure of unknown chemical species or the composition of mixtures. 

NMR spectroscopy is quantitative, chemically specific and non-invasive and therefore can be used to study 
molecules \textit{in situ} \citep{gladden1994}.   
A major limitation of conventional NMR spectroscopy is its low senstivity, which has led 
researchers to develop ever more powerful magnets to amplify the signal \citep{Barrett2013}.  However, stronger magnetic fields 
have limited returns: they are costly and also impractical in many applications, e.g., plant measurements in the chemical industry \citep{dalitz2012process}. There are 
therefore strong drivers to develop alternative techniques that can provide chemical information from relatively poor quality data.  

In this paper, we propose an alternate approach to NMR spectroscopy.  In our approach, we model the free induction decay (FID) signal as a 
realisation from a probabilistic generative model, and we use Bayesian inference and likelihood principles to infer latent variables in this
model.  This model allows us to leverage 
additional information in the FID over conventional Fourier transform spectroscopy, with explicit models of signal decay, phase shifts,
frequencies, and noise, and prior information, e.g.\, about resonant frequencies.  

In a body of pioneering work \citep{bretthorst1990, evilia1993bayesian, dou1995, andrec1998, rubtsov2007time, hutton2009, aboutanios2012instantaneous}, 
discussed further in section~\ref{sec: relatedwork}, 
various statistical models have been proposed to model the FID directly.  These models, while promising, have not yet been widely adopted,
with practicioners typically preferring conventional Fourier transform spectroscopy.  We further pursue the line of research instigated by
\citet{bretthorst1990}, to develop a statistical model of the FID which we show can be used as a highly practical tool. 
In particular, we introduce new modelling parameters (such as time delays), new 
estimation procedures, detailed stress tests, detailed comparisons with conventional methods, implementational details necessary
for good performance, and the application to chemical quantification.  In general, our approach is robust to noise, 
and allows one to study systems in which there is significant overlap between peaks in the Fourier transform of the FID signal, and
a large number of resonant frequencies.  While most approaches to NMR spectroscopy are applied in the high resolution $^1 \text{H}$ spectrum, 
we focus on examples in the $^{13} \text{C}$
spectrum, which is $10^4$ times less sensitive.

We begin with a brief introduction to conventional Fourier transform spectroscopy
in section~\ref{sec: convspec}.  We then introduce our alternative model for 
free induction decay in section~\ref{sec: modelspec}, and discuss inference 
and parameter estimation in this model in sections \ref{sec: amps} and 
\ref{sec: estothers}.  We resolve general practical difficulties
with parameter estimation in statistical models of free induction decay -- 
difficulties which have not been discussed at length in prior work.
In particular, estimating a large number of resonant frequencies, jointly
with phase parameters and other variables, requires careful navigation of
a highly multimodal likelihood surface.

In our experiments of sections \ref{sec: experiments} and \ref{sec: realexperiments}, we compare conventional
Fourier transform spectroscopy with the proposed method for predicting the 
relative concentrations of chemicals in a mixture, using both synthetic and
experimentally acquired FID signals.  We show that the proposed
model enables accurate predictions which are not possible with conventional 
Fourier transform spectroscopy.

\section{Conventional Spectroscopy}
\label{sec: convspec}

In this section we outline a conventional Fourier transform 
spectroscopy approach \citep{keeler2011understanding, malz2005}. 
Then, in the following sections, we will 
propose an alternative method for NMR spectroscopy, and 
compare both the conventional and proposed approach for
quantifying relative chemical concentrations in a mixture.

To understand how NMR spectroscopy works at a high level,
imagine a chemical mixture is placed in a strong magnetic field.  
Nuclei within this mixture that have non-zero spin will interact
with the magnetic field to produce ``magnetic moments''. 
We can imagine these magnetic moments as small bar 
magnets. Radio frequency (rf) pulses are then directed at the mixture, 
exerting a torque on these ``bar magnets'', causing them to precess perpendicular
to the strong magnetic field.  The rotating magnets create
a magnetic flux through a coil in the NMR machine, which induces 
a current with periodic components at the resonant frequencies
of the chemical mixture.  The voltage associated with this 
current is what is measured in an NMR experiment and is 
called the free induction decay (FID).  The resonant frequencies
are sensitive to the local molecular structure, which is what 
permits NMR to be used as a spectroscopic technique.

In conventional Fourier transform spectroscopy, the FID signal is assumed to 
be generated by two channels, exactly $\pi/2$ out of phase, with each channel 
perfectly modelled as a noise free sum of cosines at different frequencies 
$\omega_i^{(j)}$ weighted by intensities $B^{(j)}_i$.   
Thus taking the Fourier transform of the FID, in principle, would result in a series of spikes (delta functions) at
the resonant frequencies, and the relative magnitudes of these spikes (after adjusting for the
intensities $B^{(j)}_i$) would be the
relative concentrations of chemicals in a mixture.  However, in actuality, there is
decay in the signal due to, e.g., variations in the magnetic field \citep{keeler2011understanding} (see also section \ref{sec: modelspec} for more detail).
Moreover, the trigonometric terms are not in phase, and there is noise in the signal.  In section \ref{sec: modelspec},
we explicitly model these properties of the FID.

Throughout our experiments we will consider mixtures of cyclohexane and butanone, so we will 
exemplify the variant of conventional Fourier transform spectroscopy we use in this context; for reference,
we follow the general conventional procedure outlined in \citet{keeler2011understanding} and  \citet{malz2005}.  Table \ref{tab:excitation-profile}
gives the theoretical intensities $B^{(j)}_i$ and frequencies $\omega_i^{(j)}$ for cyclohexane and butanone\footnote{The intensity $B^{(j)}_i$ is the weighting of the $i^{\text{th}}$ cosine term, for the $i^{\text{th}}$ resonant frequency of chemical species $j$.  See, e.g., Eq.~\eqref{eqn: generative}.}.
Because the NMR spectrometer can only excite spins within a certain resonance
frequency range, the extent to which each chemical group is excited differs
from each other slightly, and this discrepancy is described by what is known as 
an \textit{excitation profile}.  This phenomenon alters the known intensities 
 $B^{(j)}_i$.
We therefore run a calibration experiment to measure the excitation profile, 
analogous to slice excitation measurements in magnetic resonance imaging \citep{haacke1999magnetic}.
We then adjust the theoretical intensity values accordingly.
The calibrated $B^{(j)}_i$ are listed in
Table~\ref{tab:excitation-profile}.  In a given experiment, the resonant frequencies will somewhat
differ from those given in Table~\ref{tab:excitation-profile}, depending on, for example, the composition of the
mixture in question.  In a conventional procedure, one must look for peaks near tabulated reference 
frequencies, and choose which chemical they are associated with.  By contrast, the proposed method in
the next sections automatically estimates the resonant frequencies for a given chemical mixture.

\begin{figure}
\centering
\subfigure[]{%
	\label{fig:fidreala}%
	\includegraphics[width=0.45\linewidth]{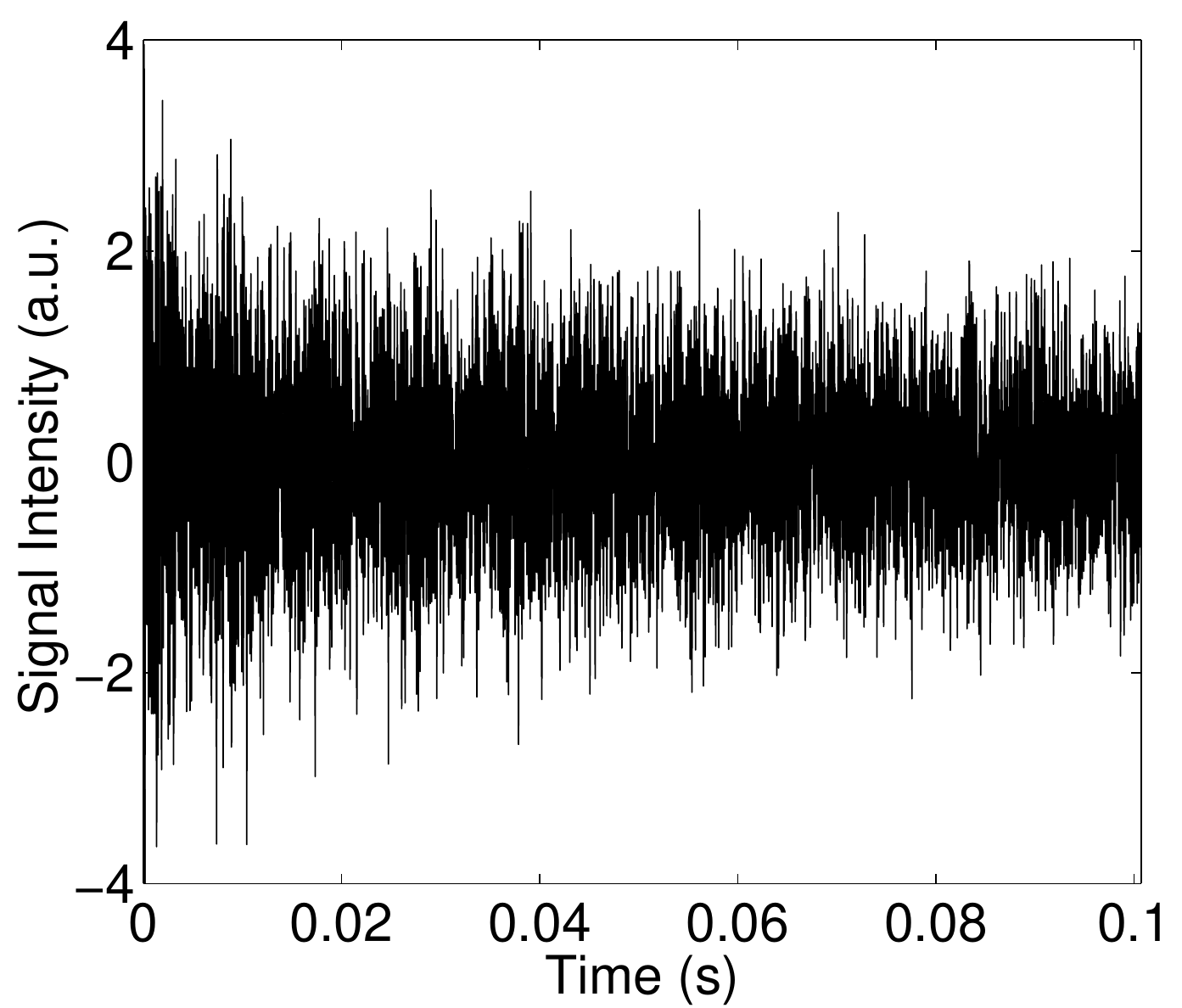}%
}%
\hspace{0.5cm}%
\subfigure[]{%
	\label{fig:transformrealb}%
	\includegraphics[width=0.45\linewidth]{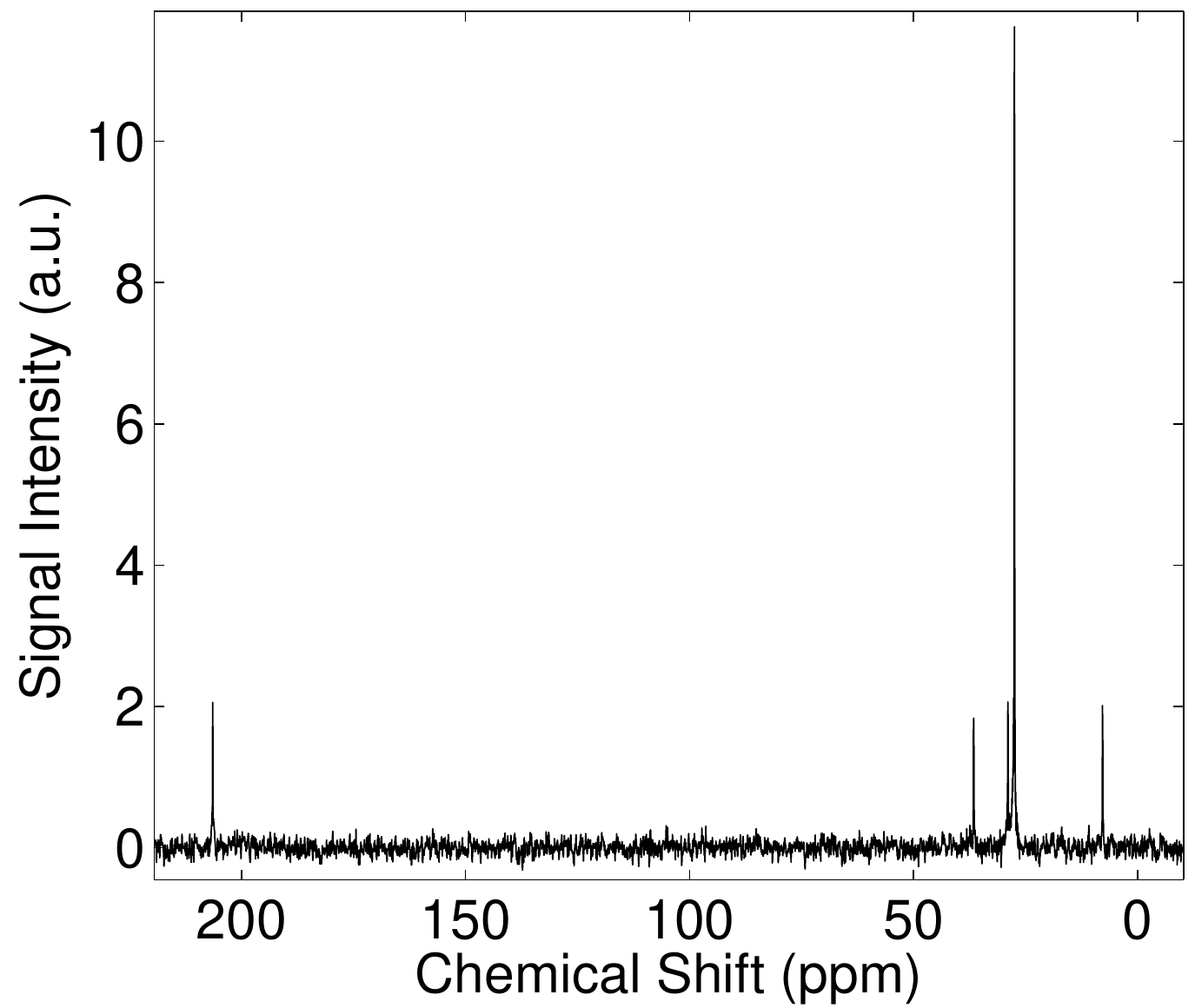}%
}%
\caption{Conventional Fourier Transform Spectroscopy for a 50-50 mixture of 2-butanone and cyclohexane.  a) Free Induction Decay (FID) Signal. b) 
Discrete Fourier Transform of FID.   }
\label{fig:convspec}
\end{figure}

Let $y(t)$ be a given FID signal, as shown in Figure \ref{fig:fidreala}. We Fourier transform $y(t)$ using the Discrete Fourier Transform (DFT) to obtain 
$\tilde{y}(\omega)$, as in Figure \ref{fig:transformrealb}.\footnote{As discussed further in section \ref{sec: convspec}, we zero-fill the data $y(t)$ prior to 
taking the Fourier transform \citep{malz2005}.} The conventional procedure is then to search for each peak near an expected resonant 
frequency.  Since the FID undergoes decay, is noisy, etc., and thus does not perfectly conform to the assumptions of a DFT,
there will be width about each peak (e.g., the peaks are not delta function spikes, but look more like Gaussian or Cauchy
densities), noise, and overlapping peaks which can be hard to differentiate in practice, particularly given that resonant 
frequencies will shift in a mixture, as seen in Figure \ref{fig:transformrealb}.

Following standard procedure, we identify a peak window for each resonant frequency and sum together the intensity of 
$\tilde{y}(\omega)$ within this peak window.  The summed intensity is then assigned as the total intensity for this 
peak window. 
Chemical quantification predictions are sensitive to the peak window, which is chosen heuristically \citep{malz2005}. 
Thereafter we are left with $K$ peaks,
belonging to chemical $j$, indexed as $S_{k}^{(j)}$.  We then calculate
\begin{align}
 I_j = \sum_{k} \frac{S_{k}^{(j)} / B_{k}^{(j)}}{K} \,,
\end{align}
where $B_{k}^{(j)}$ is the adjusted intensity in Table \ref{tab:excitation-profile}.  Using this conventional approach, 
the concentration of the $j^{\text{th}}$ chemical in the mixture is then calculated as
\begin{align}
 C_j = I_j / \sum_{j} I_j  \,.  \label{eqn: chemicaljfourier}
\end{align}
We estimate an error $E_j$ on this estimate $C_j$ using
\begin{align}
 E_j = C_j \sqrt{(\frac{E_{I_j}}{I_j})^2 + \frac{\sum_{j} E_{I_j}^2}{(\sum_{j}I_{j})^2}} \,,
\end{align}
where 
\begin{align}
 E_{I_j} = \sqrt{\sum_{k} \frac{n_{k}^{(j)} \sigma_S^2}{(K B_{k}^{(j)})^2}} \,. \label{eqn: conventionalerror}
\end{align}
$n_{k}^{(j)}$ is the number of samples along the frequency axis used to defined a peak window for the $k^{\text{th}}$ peak of
the $j^{\text{th}}$ chemical (e.g., the number of samples used to calculate $S_{k}^{(j)}$), and $\sigma_S$ is the standard
deviation of the noise in the spectral domain $\tilde{y}(s)$.

In short, using a conventional Fourier transform spectroscopy method we calculate the 
relative concentration of chemical $j$ as $C_j$ in Eq.~\eqref{eqn: chemicaljfourier}, and the uncertainty about this estimate
as $2 E_j$ in Eq.~\eqref{eqn: conventionalerror} to approximate a 95\% credible interval.

Finally, the signal to noise ratio (SNR) of the FID, which we will refer to in the experiments of sections \ref{sec: experiments} and \ref{sec: realexperiments},
is given as
\begin{align}
 \text{SNR} = \frac {S_{k}^{(j)}} {B_{k}^{(j)} \sigma_S} \,. \label{eqn: SNR}
\end{align}
Using the definition of Eq.~\eqref{eqn: SNR}, the SNR will be different for each peak $S_{k}^{(j)}$. 
When we quote the SNR, we always state the value for the lowest intensity peak.

In both simulations and real experiments, we consider the FID sampled at $4029$ points at $25 \mu s$ intervals.  For conventional 
Fourier transform spectroscopy, the time domain data were first apodised
with an exponential function and zerofilled to 16384 points to allow a sufficient
spectral resolution. 
Then, a Fourier transform was applied to the FID.  
A standard baseline correction was then performed on the spectrometer: 
we fit a polynomial to the Fourier transform of the FID (the spectrum), and subtract the polynomial fit from the spectrum
to product a flat baseline.  All such conventional processing
was performed in \textit{TopSpin} (Bruker)\footnote{\url{http://www.bruker.com/products/mr/nmr/nmr-software/software.html}}.

\begin{table}
\centering
\caption{Resonant frequencies and known intensities for each chemical group in
the mixture. The frequencies and theoretical $B^{(j)}_i$ are tabulated from
the National Institute of Advanced Industrial Science and Technology Database:
\url{http://sdbs.riodb.aist.go.jp}.}
\label{tab:excitation-profile}
\begin{tabular}{c>{\centering\arraybackslash}m{2cm}>{\centering\arraybackslash}m{2cm}>{\centering\arraybackslash}m{2cm}>{\centering\arraybackslash}m{2cm}}
\hline
Chemical
	& Frequencies (ppm)
	& Excitation profile weighting
	& Theoretical $B^{(j)}_i$
	& Calibrated $B^{(j)}_i$\\
\hline
2-Butanone & $209.29$ & $90.79\%$ & $1$ & $0.908$\\
	& $36.87$ & $96.71\%$ & $1$ & $0.967$\\
	& $29.43$ & $95.72\%$ & $1$ & $0.957$\\
	& $7.87$ & $93.03\%$ & $1$ & $0.93$\\
Cyclohexane & $27.1$ & $95.59\%$ & $6$ & $5.735$\\
\hline
\end{tabular}
\end{table}

\section{Proposed Model Specification}
\label{sec: modelspec}

We now propose an alternative to conventional Fourier transform methods for 
NMR spectroscopy.

Ultimately, we wish to predict the relative concentrations of chemicals in a
mixture, from a time-varying free induction decay (FID) signal.  Typically the
FID is recorded in two channels, $y_1(t)$ and $y_2(t)$, assumed $\pi/2$ out of
phase (e.g., real and imaginary parts of a complex signal), and independently
corrupted with i.i.d.\ Gaussian white noise.

\begin{figure}
\includegraphics{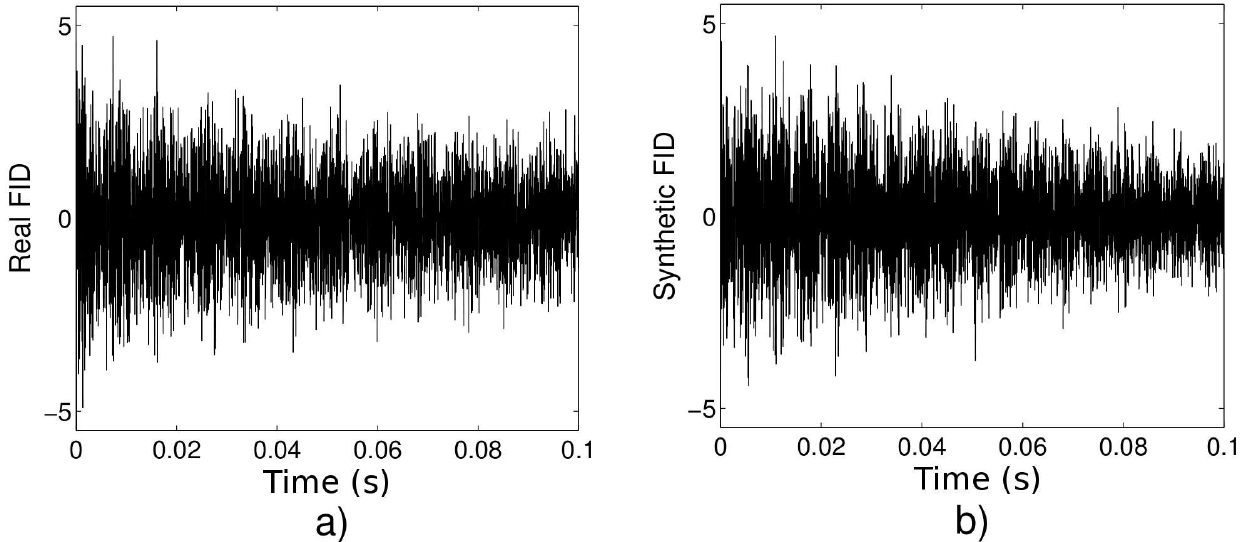}
\caption[]{a) Real (experimental) and b) Synthetic free induction decay
signals for a 30\% cyclohexane 70\% 2-butanone mixture, sampled at 25$\mu s$
intervals, from a single channel.}
\label{fig: fidcompare}
\end{figure}

Assuming $r$ chemical species in a mixture, we model $y_1(t)$ as 
\begin{align}
 y_1(t) &= A_1 \sum_{i=1}^{m_1} B_{i}^{(1)} \cos((\omega_i^{(1)} - \omega_0)(t+\tau) + \theta)e^{-\alpha t} + \dots \notag \\
      &+ A_r \sum_{i=1}^{m_r} B_{i}^{(r)} \cos((\omega_i^{(r)} - \omega_0)(t+\tau) + \theta)e^{-\alpha t} \notag \\
      &+ \epsilon_1(t) \,, \label{eqn: generative}
\end{align}                                                                                 
with amplitudes $A_1, \dots, A_r$, frequencies $\{\omega\}_i^{(j)}$ and
reference frequency $\omega_0$ in \texttt{rad/s} (chemical species $j$ has
$m_j$ resonant frequencies), intensities $\{B\}_i^{(j)}$, decay constant
$\alpha$ in \texttt{1/s}, global phase shift $\theta$ in \texttt{rad}, time
delay $\tau$ in \texttt{s}, and noise $\epsilon_1(t) \sim \mathcal{N}(0,v)$.\footnote{To convert from the \texttt{ppm} of Table \ref{tab:excitation-profile}
into \texttt{rad/s}, we use \text{ppm} $\times$ $75$ \texttt{Hz} (for the $^{13}$C spectrum on our $7$T magnet)
$\times$ $2 \pi$.}  We will always assume that the intensities $\{B\}_i^{(j)}$ and reference
frequency $\omega_0$ are known and fixed, and we will estimate all other 
parameters in a procedure outlined in sections \ref{sec: amps} and \ref{sec: estothers}. In Figure~\ref{fig: fidcompare} we
show a draw from this generative model, compared with a real FID signal, for a
30\% cyclohexane 70\% 2-butanone mixture.  The real and synthetic FID signals
are qualitatively similar -- similar resonant frequencies, similar
decay rates, etc.  

The parameters in the generative model of Eq.~\eqref{eqn: generative} have clear physical interpretations.
The excitation of a sample, e.g. a chemical mixture, in NMR is provided by a radiofrequency (rf) coil which resonates at the frequency of the 
nucleus of interest, e.g. $^{13}$C.  The time delay $\tau$ in our model arises from the `ringdown' time of this 
rf coil. The ringdown time is the time taken for the current in the coil to cease. In NMR, rf coils are 
designed for maximum sensitivity and therefore resonate for a prolonged period after the current supplied 
to the coil is stopped.  The signal from the sample cannot be detected until this ringdown period has finished. 
In a typical coil this time will be on the order of a few microseconds.  In conventional Fourier transform 
spectroscopy, this delay between excitation and starting the signal detection leads to a linear change in the 
phase of the signal with frequency variation, which is corrected after Fourier transformation of the raw signal.
In our system, we include the time delay in the model directly.

The signal lifetime in NMR is limited by both the relaxation and homogeneity of the magnetic field in which the 
sample is placed. Both of these effects lead to the decay $\alpha$ of the signal amplitude over time.  In an ideal system, 
the signal will decay through an exponential process \citep{keeler2011understanding}. Perhaps surprisingly, this model is also appropriate to 
a range of non-ideal systems including, e.g., the signal from liquids in rocks \citep{chen2005internal}. 
Here we study pure liquids and so the exponential decay model is most 
appropriate, though we note that our methodology could easily be adapted for a variety of other models describing the signal decay.

We can rewrite Eq.~\eqref{eqn: generative} as an inner product of amplitude
coefficients $\bm{a} = (A_1,\dots,A_r)^{\top}$ with cosine basis functions
$\bm{\phi} = (\phi_1,\dots,\phi_r)$ parametrized by $\psi = \{
\{\omega\}_i^{j}, \theta, \tau, \alpha \}$:
\begin{align}
 y_1(t) &= \bm{a}^{\top} \bm{\phi}(t,\psi) + \epsilon_1(t) \,, \\   \label{eqn: realfid}
 \epsilon_1(t) &\sim \mathcal{N}(0,v) \,.
\end{align}
The model for $y_2(t)$ is the same, 
\begin{align}
 y_2(t) &= \bm{a}^{\top} \bm{\varphi}(t,\psi) + \epsilon_2(t) \,, \label{eqn: imfid} \\
 \epsilon_2(t) &\sim \mathcal{N}(0,v) \,,
\end{align}
except we use sine instead of cosine basis functions in $\bm{\varphi}$ to
account for the $\pi/2$ phase difference between the two channels.

In section~\ref{sec: amps} we infer $p(\bm{a}|\mathcal{D},\{t_n\}_{n=1}^{N},
\psi, v)$, a posterior distribution over the amplitude variables given an
observed FID signal $\mathcal{D}$ at times $\{t_n\}_{n=1}^{N}$, and nuisance
variables $\psi$ and $v$.  Since the relative concentrations of chemicals $i$
and $j$ is given by $|A_i/A_j|$, this posterior distribution can be used to
estimate, with uncertainty intervals, the relative concentrations of chemicals
in a mixture.

In section~\ref{sec: estothers} we discuss estimation of frequency variables
$\{\omega\}_i^{(j)}$, other nuisance variables in $\psi$, and the noise
variance $v$.  Performance in estimating chemical concentrations is especially
sensitive to frequency estimates.  Moreover, accurately estimating frequency
parameters is difficult because the likelihood of the data is highly multimodal
as a function of these frequency variables.

Incidentally, it may be a slight misnomer to refer to $\psi, v$ as nuisance
variables.  Although we are primarily interested in estimating the relative
concentrations of chemicals in a mixture -- and thus, the $\bm{a}$ variable
-- the learned frequencies, decay
rates, phase shifts, and noise variance are all still enlightening.  For instance, 
the learned frequencies could be used, in principle, to
better understand how resonant frequencies shift for particular chemicals when
immersed in a mixture.

\section{Inference over Amplitudes}
\label{sec: amps}
We wish to infer $p(\bm{a}|\bm{y}_1,\bm{y}_2,\psi,v)$, where $\bm{y}_1$ and
$\bm{y}_2$ are the ``real'' and ``imaginary'' channels respectively observed
at times $\{t_{1n}\}_{n=1}^{N}$ and $\{t_{2m}\}_{m=1}^M$.  To reflect our 
prior uncertainty over the amplitudes, we specify a prior $p(\bm{a}) = \mathcal{N}(\mu_0,S_0)$, and find
\begin{align}
p(\bm{a}|\bm{y}_1,\bm{y}_2,\psi,v) = \mathcal{N}(\mu,S) \,, \label{eqn: aposterior} \\
S = (S_0^{-1} + \frac{1}{v}\Phi^{\top}\Phi + \frac{1}{v}\Psi^{\top}\Psi)^{-1} \,, \label{eqn: S} \\
\mu =  S(S_0^{-1}\mu_0 + \frac{1}{v}\Phi^{\top}\bm{y}_1 + \frac{1}{v}\Psi^{\top}\bm{y}_2) \label{eqn: mu} \,,
\end{align}
where 
$\Phi$ and $\Psi$ are respectively $N \times r$ and $M \times r$ matrices with entries
\begin{align}
\Phi_{i,j}(\psi,t_i) =  \bm{\phi}_j(\psi,t_{1i}) \,, \\ 
\Psi_{i,j}(\psi,t_i) =  \bm{\varphi}_j(\psi,t_{2i}) \,.
\end{align}
We may not know anything about the relative concentrations of the chemicals a priori, 
in which case we wish to have a vague uninformative prior over the amplitudes.  For 
example, one can use an improper uniform prior on the amplitudes by taking
$S_0 = \lim_{\gamma \to \infty} \gamma I$, which causes the $S_0^{-1}$ and
$S_0^{-1}\mu_0$ terms to vanish in Equations \eqref{eqn: S} and \eqref{eqn:
mu}, respectively.  Alternatively, one can start with a vague (much higher variance
than mean) prior (effectively a uniform prior), and then learn the value of $\gamma$
from the data following the procedure outlined in section \ref{sec: estothers}.  

The posterior distribution over the amplitudes is
conditioned on the nuisance variables $\psi$ and $v$. Before we can use this
posterior to make predictions about the relative concentrations of chemicals
in a mixture, we must first either integrate away these nuisance variables, or
find a point estimate, conditioned on the data.  

\section{Estimating Nuisance Variables}
\label{sec: estothers}

We find that the performance of the model in Equations~\eqref{eqn: realfid}
and~\eqref{eqn: imfid} is sensitive to estimates of the nuisance variables,
particularly the frequency variables.  Furthermore, estimating these
frequencies is non-trivial, due to a severely multimodal likelihood.

As a step towards estimating the nuisance variables, $\psi$ and $v$, we can
analytically integrate away the amplitude variables from the \text{likelihood}
of the data $p(\bm{y}_1,\bm{y}_2 | \bm{a}, \psi, v)$:
\begin{align}
 p(\bm{y}_1,\bm{y}_2 | \psi, v) &= \int p(\bm{y}_1,\bm{y}_2,\bm{a} | \psi, v) d\bm{a} \label{eqn: firstml} \\
  &= \int p(\bm{y}_1,\bm{y}_2 | \psi, v, \bm{a}) p(\bm{a}) d\bm{a} \label{eqn: secondml} \\ 
  &= \int p(\bm{y}_1 | \bm{a}^{\top} \bm{\phi}(t,\psi), v)  p(\bm{y}_2|\bm{a}^{\top} \bm{\varphi}(t,\psi), v) p(\bm{a}) d\bm{a} \label{eqn: thirdml} \\
  &= \int p(\bm{a} | \bm{y}_1, \psi, v) p(\bm{y}_1 | \psi,v) p(\bm{y}_2|\bm{a}^{\top} \bm{\varphi}(t,\psi), v) d\bm{a} \label{eqn: fourthml} \,.
\end{align}
In Equation~\eqref{eqn: thirdml}, the likelihood of the data decomposes into a
product of Gaussian likelihoods for each channel, since the channels are
independent given the noise free signal.  Performing the integration in
Eq.~\eqref{eqn: fourthml}, assuming $S_0 = \gamma I$ for notational
simplicity, we find the log \textit{marginal} likelihood of the data is
\begin{align}
\log p(\bm{y}_1,\bm{y}_2 | \psi, v) = -\frac{1}{2v}(||\bm{y}_1 - \Phi \mu_{\text{real}}||^2 + ||\bm{y}_2 - \Psi \mu ||^2) - \frac{1}{2\gamma}||\mu-\mu_0||^2  \notag \\  
- \frac{1}{2}(\mu-\mu_{\text{real}})^{\top}(\frac{\Phi^{\top}\Phi}{v} + \frac{I}{\gamma})(\mu-\mu_{\text{real}}) - \frac{N+M}{2} \log(v) -\frac{r}{2} \log (\gamma) - \frac{N+M}{2}\log(2 \pi) \,, \label{eqn: marglike} 
\end{align}
where $\mu_\text{real}$ is the posterior mean on the amplitudes if we were to
only have data from the real channel:
\begin{align}
 \mu_\text{real} = (\frac{\Phi^{\top}\Phi}{v} + \frac{I}{\gamma})^{-1}(\frac{\mu_0}{\gamma} + \Phi^{\top} \bm{y}_1)\,.
\end{align}
The first two terms of Eq.~\eqref{eqn: marglike} are model fit terms, since
$\Phi \mu_{\text{real}}$ and $\Psi \mu$ are estimates of the real and
imaginary channels of the noise free FID.  The remaining terms are
normalization constants, and automatically calibrated complexity penalties
\citep{rasmussen01}, which come from integrating away the amplitudes from the
likelihood.

We can estimate the nuisance variables $\psi,v$ by finding the $\hat{\psi},
\hat{v}$ that maximize the log marginal likelihood $\log p(\bm{y}_1,\bm{y}_2 |
\psi, v)$ in Eq. \eqref{eqn: marglike}, a procedure sometimes called
\textit{empirical Bayes} or \textit{type-II maximum likelihood}, or we can use
the marginal likelihood to integrate away $\psi,v$ via Markov chain Monte Carlo (MCMC), 
in order to sample from the posterior $p(\bm{a}|\bm{y_1},\bm{y_2})$.  In either case, we must navigate a
challenging multimodal likelihood surface.

\begin{figure}
\centering
\subfigure[]{%
	\label{fig:3070-a}%
	\includegraphics[width=0.3\linewidth]{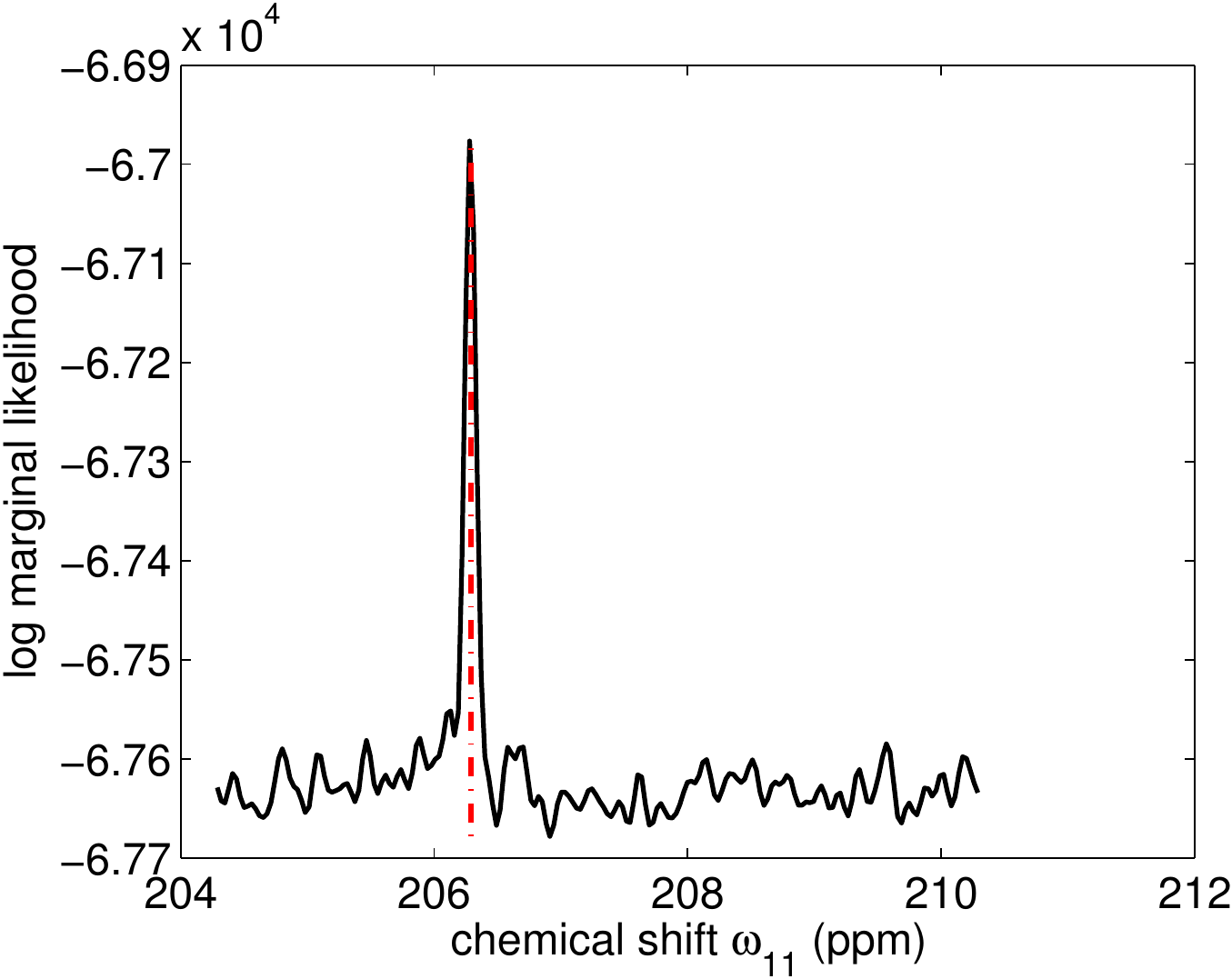}%
}%
\hspace{0.4cm}%
\subfigure[]{%
	\label{fig:3070-b}%
	\includegraphics[width=0.3\linewidth]{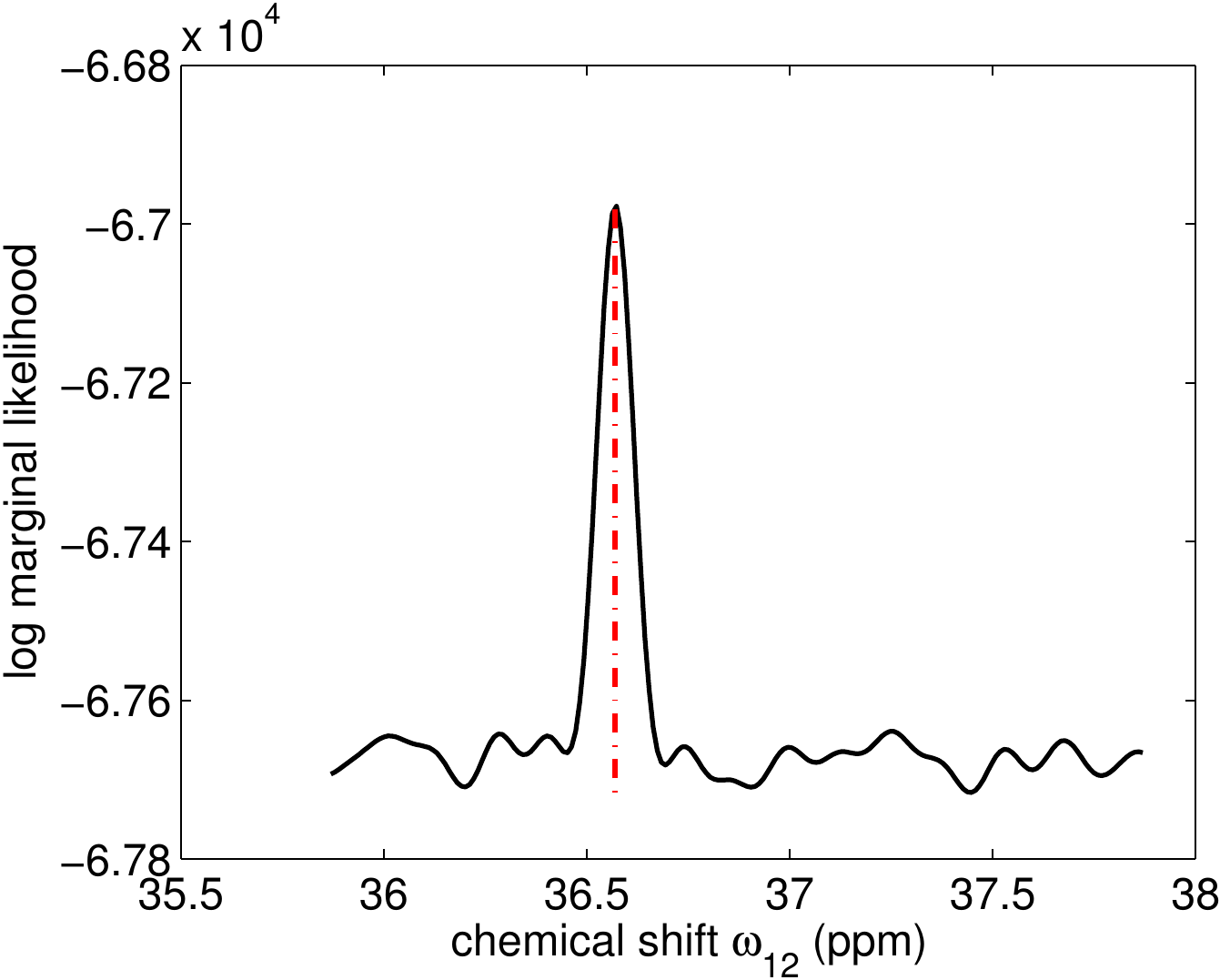}%
}%
\hspace{0.4cm}%
\subfigure[]{%
	\label{fig:3070-c}%
	\includegraphics[width=0.3\linewidth]{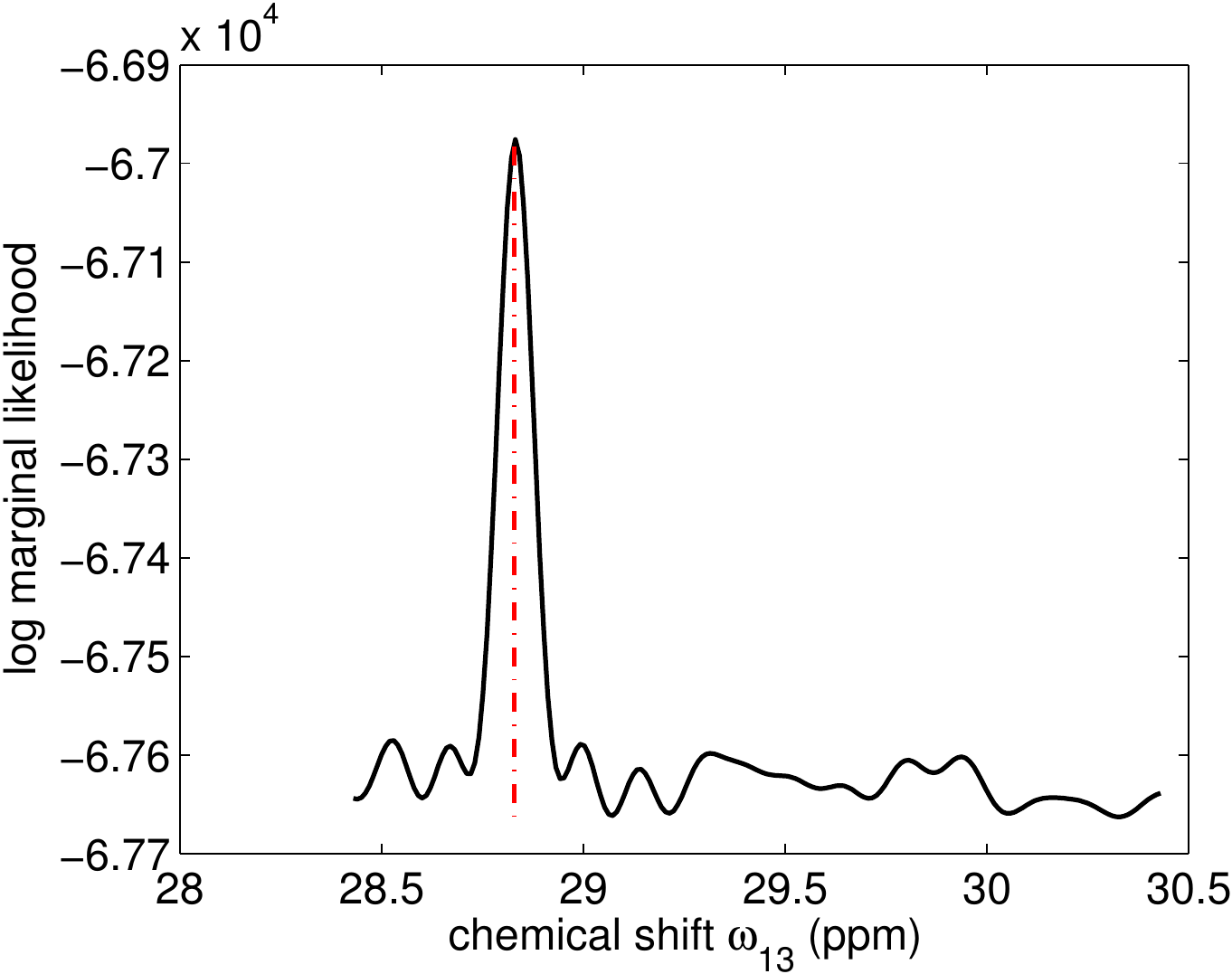}%
}%
\\
\subfigure[]{%
	\label{fig:3070-d}%
	\includegraphics[width=0.3\linewidth]{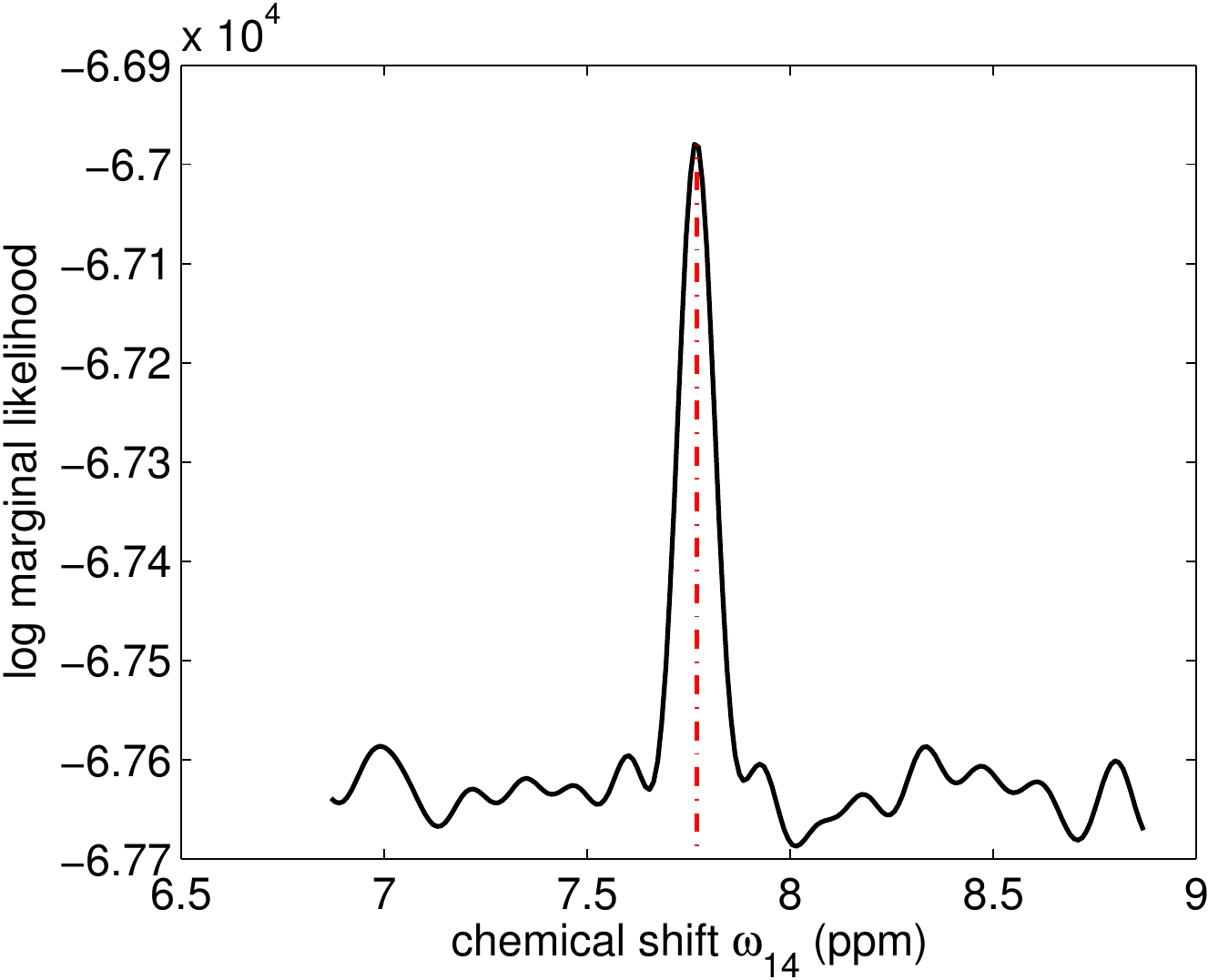}%
}%
\hspace{0.4cm}%
\subfigure[]{%
	\label{fig:3070-e}%
	\includegraphics[width=0.3\linewidth]{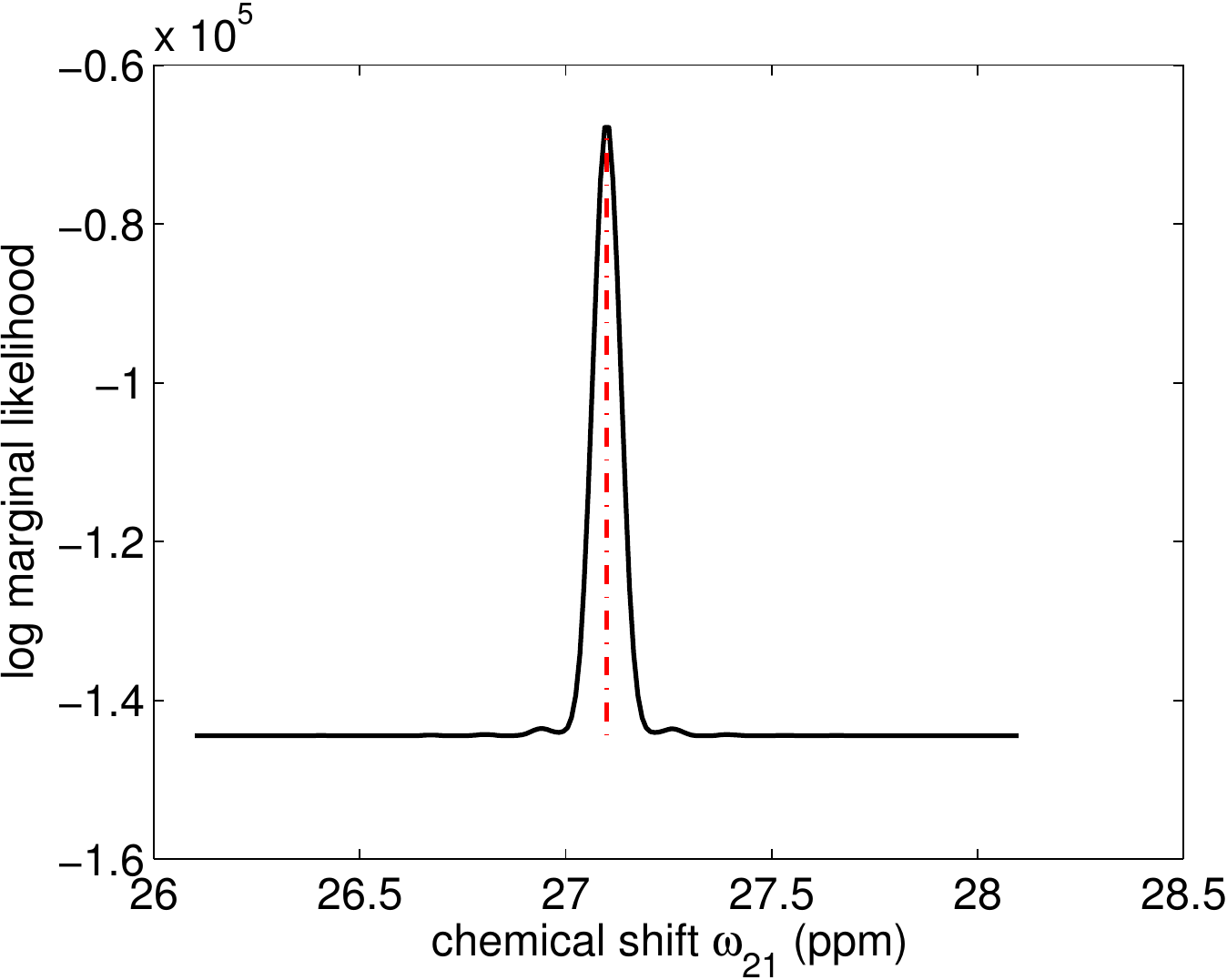}%
}%
\hspace{0.4cm}%
\subfigure[]{%
	\label{fig:3070-f}%
	\includegraphics[width=0.3\linewidth]{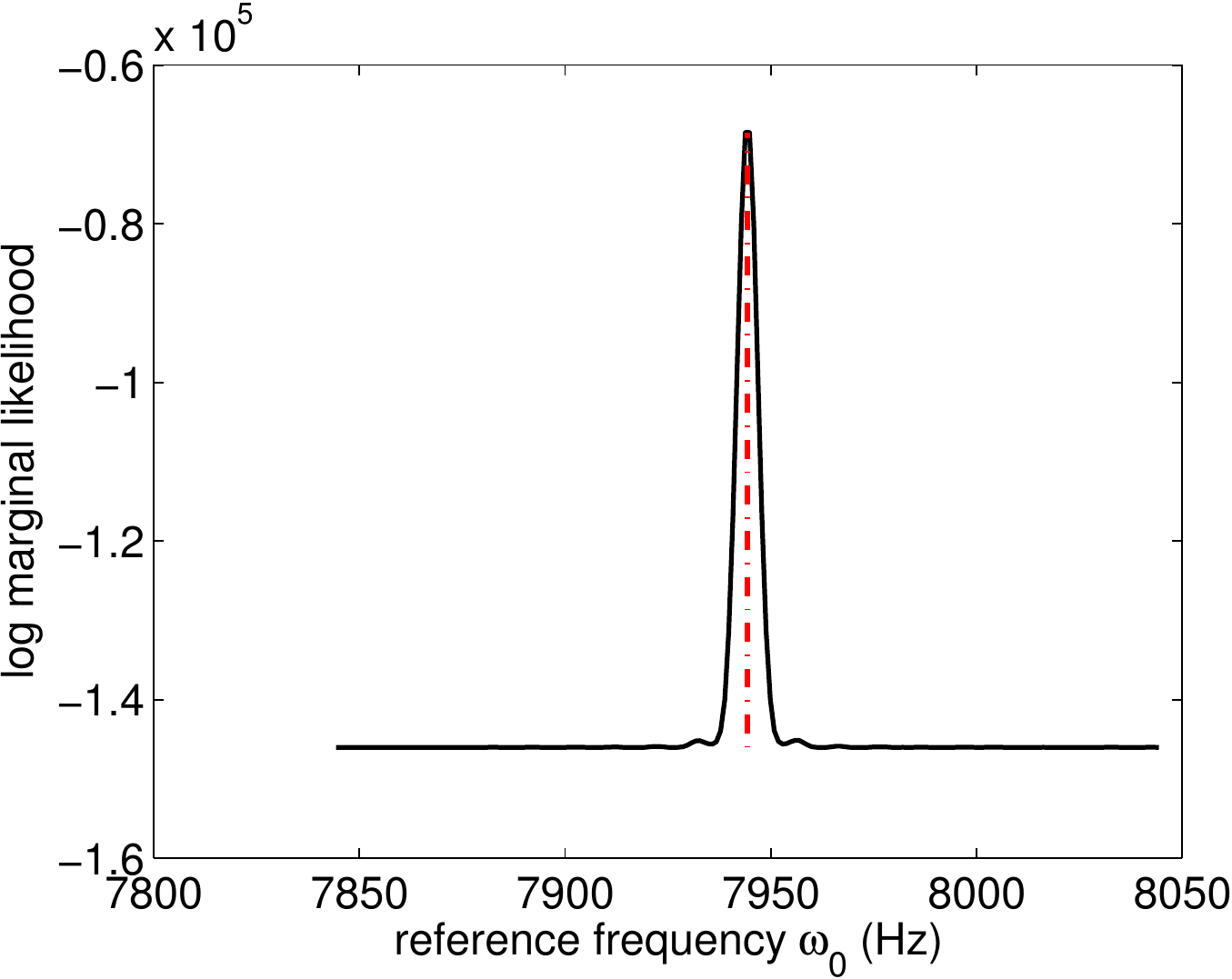}%
}%
\\
\subfigure[]{%
	\label{fig:3070-g}%
	\includegraphics[width=0.3\linewidth]{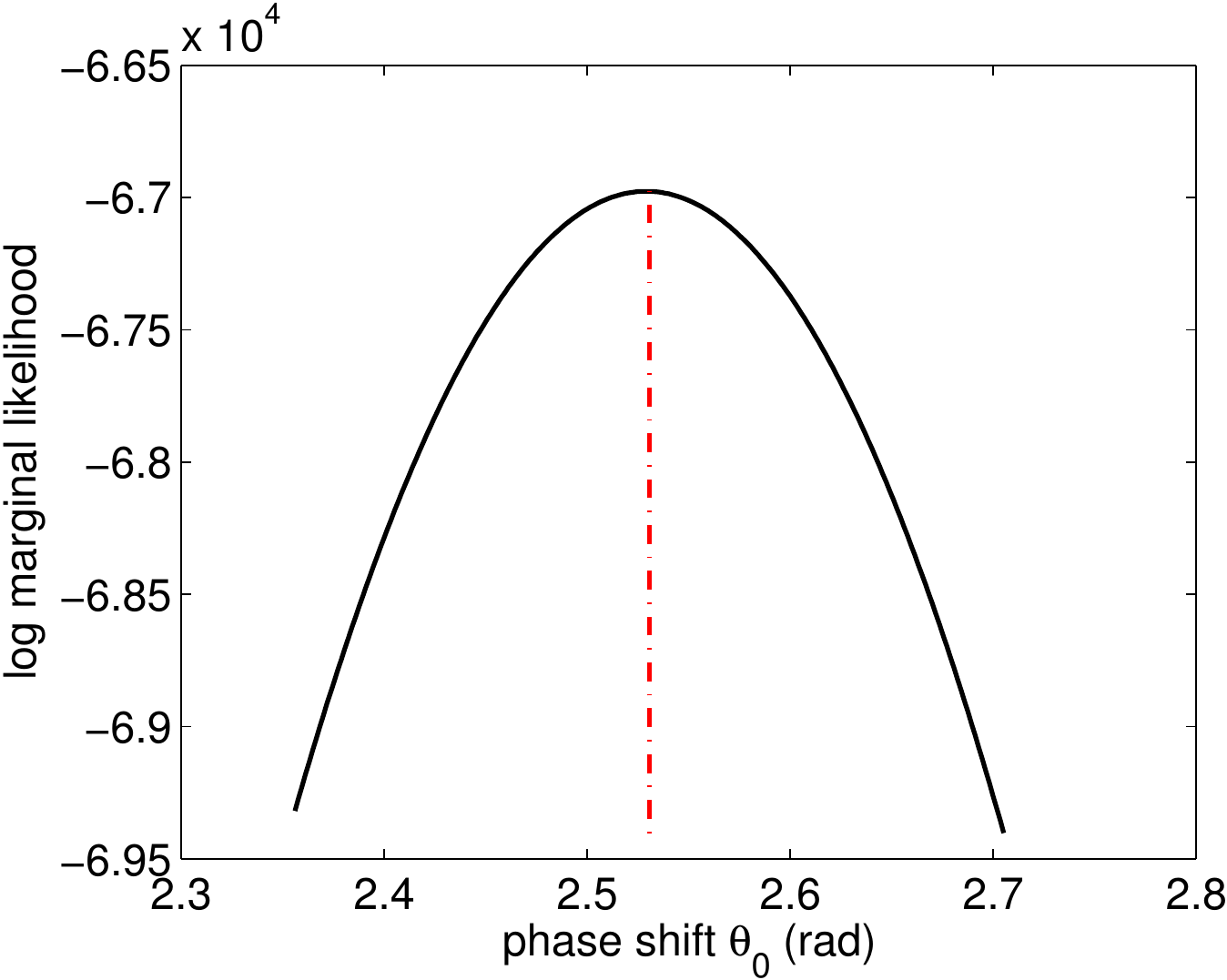}%
}%
\hspace{0.4cm}%
\subfigure[]{%
	\label{fig:3070-h}%
	\includegraphics[width=0.3\linewidth]{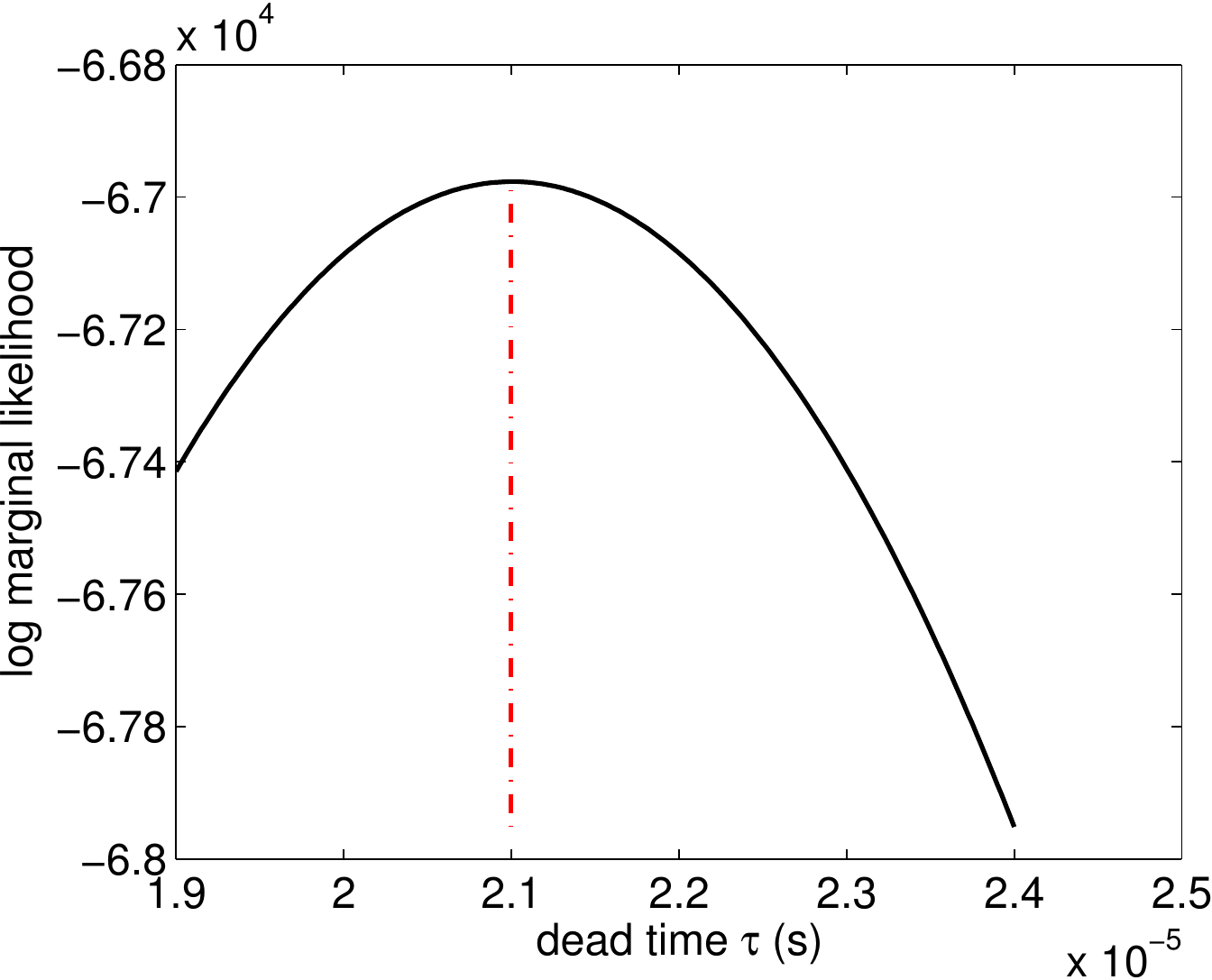}%
}%
\hspace{0.4cm}%
\subfigure[]{%
	\label{fig:3070-i}%
	\includegraphics[width=0.3\linewidth]{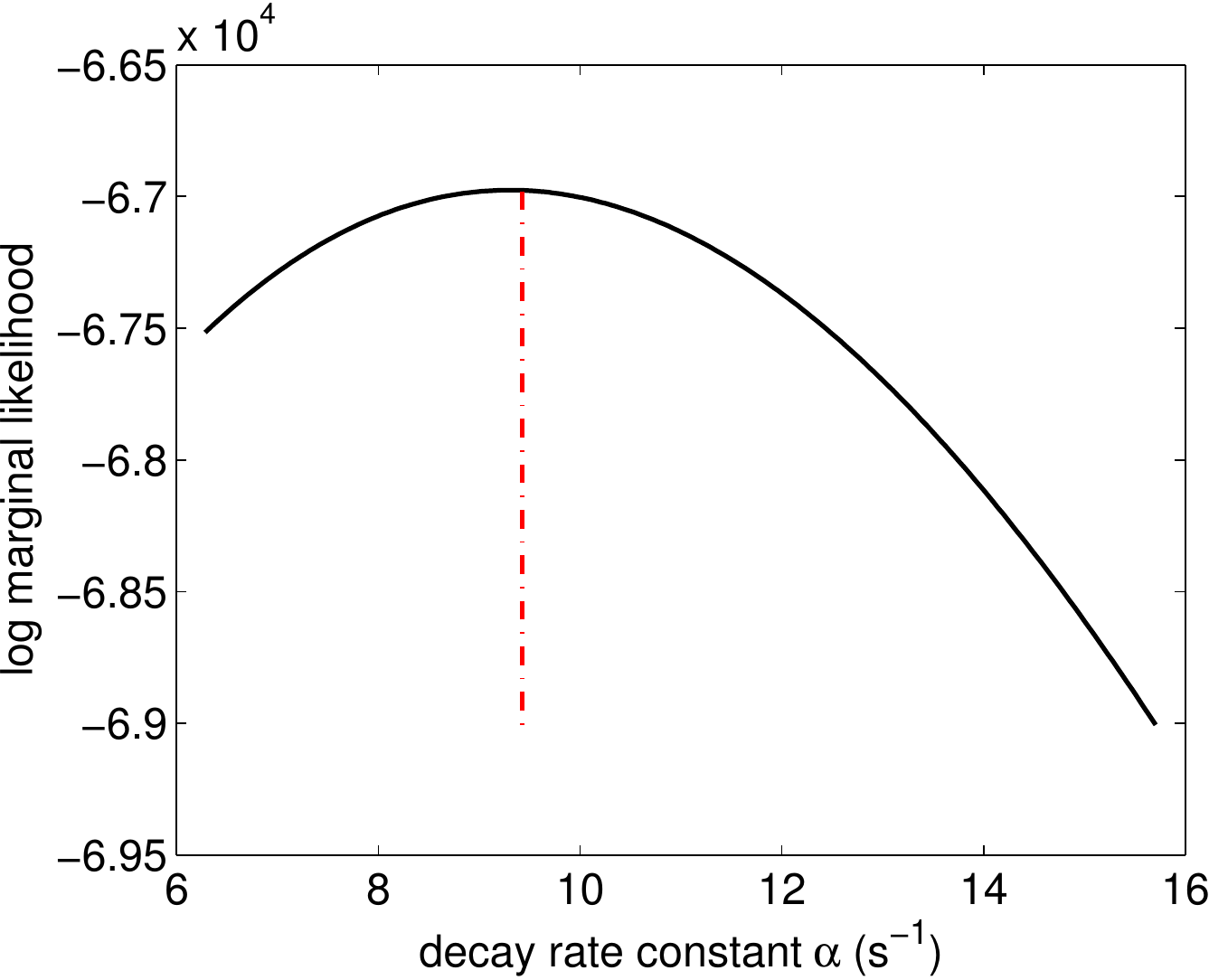}%
}%
\\
\subfigure[]{%
	\label{fig:3070-j}%
	\includegraphics[width=0.3\linewidth]{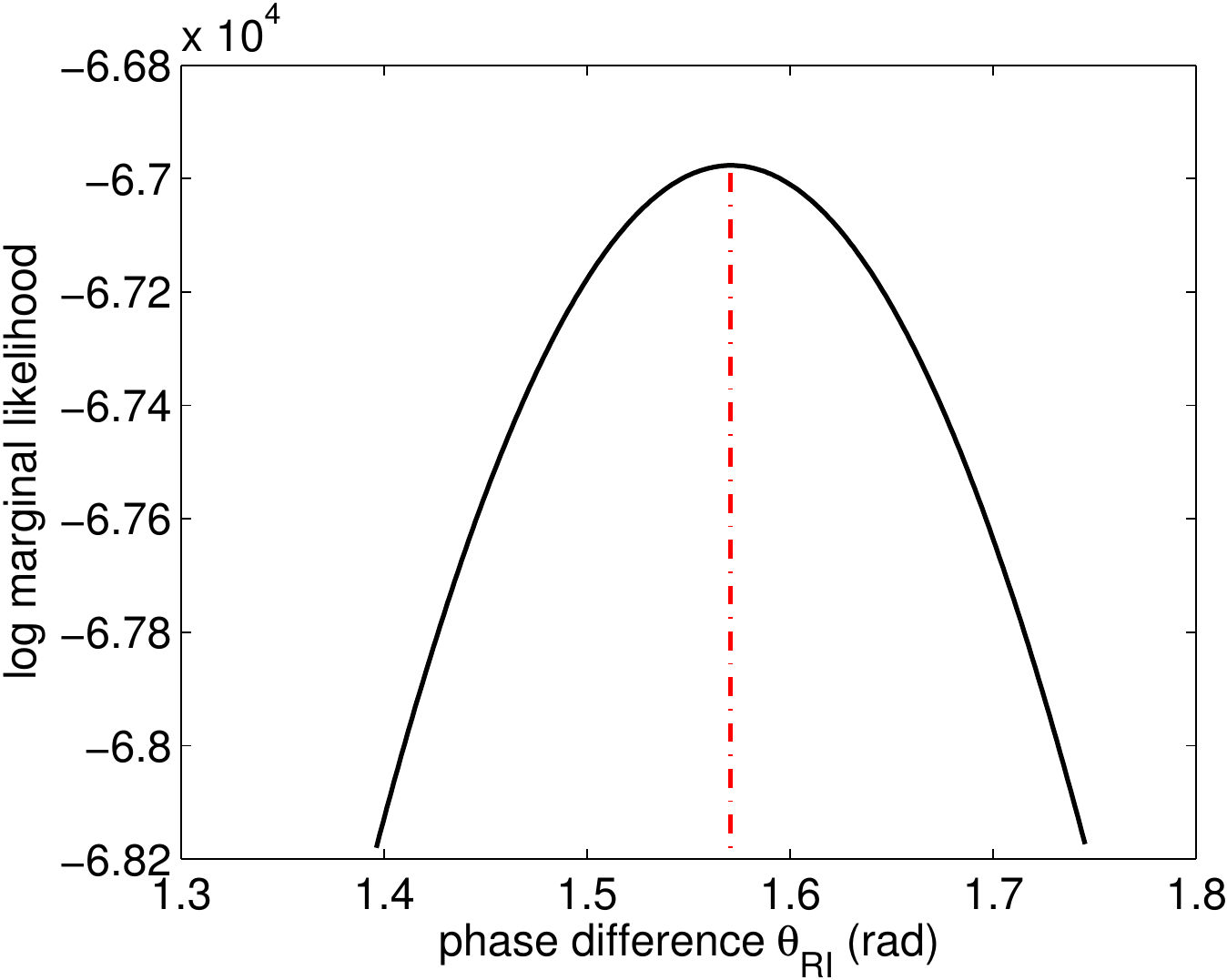}%
}%
\hspace{0.4cm}%
\subfigure[]{%
	\label{fig:3070-k}%
	\includegraphics[width=0.3\linewidth]{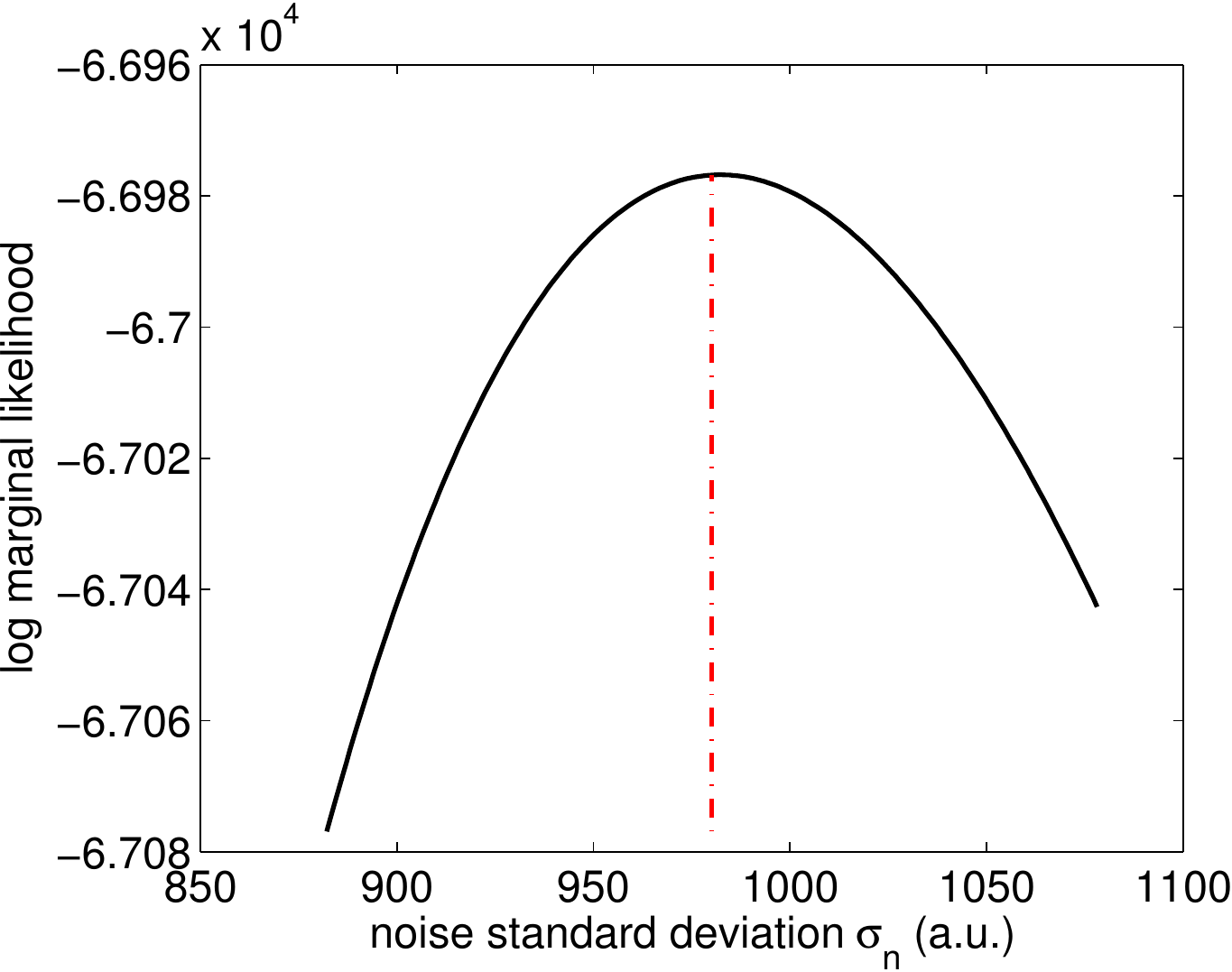}%
}%
\caption{Log marginal likelihood plots as a function of all the nuisance
parameters in simulation.  Each panel shows how the log marginal likelihood
varies with a given variable, with all other variables set to their true values.
The ground truth frequencies 
are $206.29, 36.57, 28.43, 7.77$ ppm for butanone and $27.1$ ppm for cyclohexane.
The dashed line gives the corresponding true value for the parameter of interest. 
The FID data were generated from 2-butanone ($30\%$) and cyclohexane ($70\%$) mixture
with $8$ scans, corresponding to an SNR (as defined in Eq.~\eqref{eqn: SNR} of section \ref{sec: convspec}) of 
$26$ for 2-butanone and $381$ for cyclohexane).
}
\label{fig:marginal-likelihood-vis}
\end{figure}

In Figure~\ref{fig:marginal-likelihood-vis}, we illustrate the behaviour of
the marginal likelihood in Eq.~\eqref{eqn: marglike} as a function of the
nuisance parameters $\psi, v$, on a synthetic equal mixture of cyclohexane and
2-butanone.  We generate the data by sampling from the generative model in
Eqs.~\eqref{eqn: realfid}--\eqref{eqn: imfid} with frequencies and intensities
for cyclohexane and 2-butanone as given in Table~\ref{tab:excitation-profile}.

Each panel in Figure~\ref{fig:marginal-likelihood-vis} shows how the marginal likelihood
varies as a function of a single parameter in the model
of Eq.~\eqref{eqn: generative}, with all other parameters set to the true
generating values.  The true values are shown with dashed vertical lines. Even in
simulation, the likelihood surface is severely multimodal as a function of the
frequency parameters.
Moreover, there are strong dependencies between the frequency and phase
parameters.  Therefore exploring this likelihood surface is non-trivial.  Our prior knowledge of
the frequencies is only accurate to within about $1000$ rad/s (or $2$ ppm).
We can see in Figure~\ref{fig:3070-a}--\subref{fig:3070-f} that this range
spans multiple local optima.  A naive gradient based optimizer -- greedily
choosing to improve the value of an objective function on every iteration 
-- will converge to
undesirable local optima, rendering predictions of the amplitudes unusable.
Likewise, popular sampling schemes, such as Metropolis Hastings, become
stuck in a local optimum if using a small step size, and proposals with a
large step size are rejected.   We can also see that sampling may have limited
value in this application, since the global optima in the frequency space are
sharply peaked.  Moreover, while the posterior over the amplitude parameters
is sensitive to the estimated values of the frequency parameters, this
posterior is not so sensitive to noise variance or decay that exploring the
unimodal distributions shown in Figure~\ref{fig:marginal-likelihood-vis}
would provide major performance gains.

The local optima arise because a signal composed of a sum of frequency
components can be extremely well modelled in parts, while missing the rest of
the signal entirely, by a range of erroneous frequencies.  The smaller the
signal, the less pronounced the multimodality.  Increasing an estimate on the
initial decay rate $\alpha$ decreases the amount of available information
(since the signal is then assumed to decay away quickly), and therefore
decreases the multimodality in the likelihood.  Thus increasing the estimate
of $\alpha$ has an annealing effect on the likelihood surface.  
Increasing the noise
variance has a similar effect: the more noise, the less available signal.
Indeed there are natural dependencies between the decay rate and noise
variance parameters.  An overly high decay rate means a large portion of the
signal can only be explained as noise.  In short, systematically
overestimating the decay rate, as part of a gradient based optimization
procedure, can be used to help locate a global optimum.  However, we find an
explicit simulated annealing simplex algorithm (SIMPSA)
\citep{cardoso1996,nelder1965}, which was especially designed for continuous,
bounded, global optimization, more reliable for this NMR spectroscopy
application.

In short, to estimate relative chemical concentrations, we follow a three step
procedure:
\begin{enumerate}
\item Use SIMPSA to find the $\psi_{\text{est}}, v_{\text{est}}$ that optimize
the marginal likelihood $\log p(\bm{y}_1,\bm{y}_2 | \psi, v)$ in
Equation~\eqref{eqn: marglike}, using a vague uninformative prior 
$p(\bm{a})$ on the amplitudes.
\item Analytically infer the posterior
$p(\bm{a}|\bm{y}_1,\bm{y}_2,\psi_{\text{est}},v_{\text{est}})$ as in
Equation~\eqref{eqn: aposterior}, conditioned on the maximum marginal
likelihood estimates of the nuisance parameters $\psi, v$.
\item By sampling from
$p(\bm{a}|\bm{y}_1,\bm{y}_2,\psi_{\text{est}},v_{\text{est}})$, sample
from the distribution over the relative concentrations $r_{ij} = |A_i/A_j|$ of
chemicals $i$ and $j$, $p(r_{ij} |
\bm{y}_1,\bm{y}_2,\psi_{\text{est}},v_{\text{est}})$, where $A_i$ is the
$i^{\text{th}}$ component of $\bm{a}$.
\end{enumerate}

The computational demands of the proposed procedure are dominated by the need
to evaluate the marginal likelihood of Eq.~\eqref{eqn: marglike} for a variety of
settings of $\psi$ and $v$.  Each such evaluation costs $\mathcal{O}((N+M)r^2
+ r^3)$ operations for $N$ and $M$ points respectively in the real and imaginary channels,
and $r$ chemical species: it takes $(N+M)r^2$ operations to compute
$\Phi^{\top}\Phi + \Psi^{\top}\Psi$, and $\mathcal{O}(r^3)$ operations to take the Cholesky
decomposition of this term (for solving linear systems).  Thus for a system with
a fixed number of chemicals, the required computational operations scale linearly
with the number of collected datapoints.

\section{Related Work}
\label{sec: relatedwork}

There is a body of groundbreaking work using models motivated for NMR
spectroscopy, similar in form to Equation \eqref{eqn: generative}.  
In this section we briefly describe a selection of this work, and 
some limitations.  \citet{yoon2006deterministic} provide a general review of statistical models 
with various NMR applications.

\citet{bretthorst1990} was an early pioneer of Bayesian estimation of a 
quadrature model, similar to Eq.~\eqref{eqn: generative}, but without 
the local phase shift $\tau$.  While inference over the amplitude variables
follows a similar procedure to that described in section~\ref{sec: amps},
inference over the frequency variables involves several approximations,
including a quadratic Taylor series expansion, leading to a Gaussian posterior
over frequencies.  However, in general, the posterior over frequencies can be
highly non-Gaussian (as shown in Figure \ref{fig:marginal-likelihood-vis}).  Generally, 
these approximations are not robust to a multimodal likelihood surface, and tend to
break down when there are more than a few resonant frequencies to estimate.   
\citet{evilia1993bayesian} contains a similar early quadrature NMR model, but the exact
model specification and estimation procedures are unclear, and the model is tested
on systems with at most 2 resonant frequencies. 

\citet{andrec1998} use Metropolis Hastings to sample from the posterior
distribution over parameters in a quadrature model similar to
\citet{bretthorst1990}, ultimately to estimate coupling constants in antiphase
doublets.  \citet{andrec1998} do not estimate more than two frequency
parameters at a time, and do not compare with conventional spectroscopy.
Furthermore, Metropolis Hastings (MH) is not generally suitable for exploring
multimodal likelihood surfaces, as discussed in section \ref{sec: estothers}.  In short, using an
MH proposal distribution with a small width will cause the sampler to become
trapped in undesirable local optima, while a large width proposal is extremely
unlikely to find a global optimum, and if it does, it will never move from a
point estimate, which defeats the purpose of sampling.

\citet{dou1995} perform Gibbs sampling \citep{geman1984} for parameter estimation, with similar
performance to \citet{bretthorst1990}.  Gibbs sampling involves alternately
sampling one parameter while conditioning on the others, in a cycle. 
When
there are strong dependencies between parameters, such as frequency and phase
parameters, and multimodality, Gibbs sampling is known to mix poorly \citep{murray-adams-2010a}.
Therefore a Gibbs sampling procedure may struggle in many NMR applications.

\citet{rubtsov2007time} focus on estimating the number of components (e.g., resonant frequencies)
in a model similar to \citet{bretthorst1990}, using reversible jump Markov chain Monte Carlo (MCMC).
While the model is applied to experimentally acquired data, the experiments are taken in the 
$^1 \text{H}$ spectrum, where the signal to noise ratio is much higher than in the $^{13} \text{C}$ 
spectrum.

In recent work, \citet{hutton2009} develop a specialised model for mixtures
of water and metabolites.  The model has a similar form to
\citet{bretthorst1990}, except the amplitude coefficient for water is
time-varying.  Estimation of frequency parameters follows a simulated
annealing MCMC approach.  It is difficult to determine whether the methodology
in \citet{hutton2009} is generally applicable to quadrature NMR models, since
the paper is focused on a specialised application in the ($^1$H) 
proton spectrum, where the SNR is much higher than in the
carbon spectrum (considered in our paper), there is no comparison to
conventional spectroscopy, and there is no motivation or description of the
applied sampling scheme, except that the computational costs can be
considerable.

Rather than construct a model in the time domain, \citet{astle2012bayesian} 
consider modelling preprocessed data (apodised, phase corrected, and with a 
baseline correction) in the frequency domain, leveraging positivity requirements
in this domain.  While promising, the model relies on conventional preprocessing
and implicit assumptions of the discrete Fourier transform (DFT), and thus may not 
be as general as time domain models.

Furthermore, \citet{rubtsov2010application} develop a promising statistical procedure for estimating
resonant frequencies in the $^1 \text{H}$ spectrum, independently from quadrature NMR models, and compare to 
binning.  And \citet{aboutanios2012instantaneous} develop heuristics for estimating components
in a quadrature NMR model, which show some robustness to synthetic noise; however, the dataset
considered is synthetic and has a well defined discrete Fourier transform.

Our paper focuses on aspects of quadrature NMR modelling which have not been thoroughly explored in prior work:
1) we develop a statistical model for NMR spectroscopy with a novel chemical quantification application, 2) we discuss
challenges in estimating quadrature NMR models, such as a multimodality, and propose solutions, 3) we provide  
thorough quantitative comparisons with conventional Fourier transform spectroscopy, and 4) we consider both synthetic 
and experimentally acquired data in the $^{13}$C spectrum, with low SNR, and many resonant frequencies.
 
\section{Simulations: 2-Butanone Cyclohexane Mixture}
\label{sec: experiments}

To better understand the behaviour of the proposed model (sections \ref{sec: modelspec}-\ref{sec: estothers}),
and conventional Fourier transform (FT) spectroscopy (section \ref{sec: convspec}), we first predict the relative 
concentrations of 2-butanone and cyclohexane in mixtures from synthetic free induction decay (FID) 
signals -- a controlled environment with a ground truth.\footnote{For conventional spectroscopy, we use
the known true phase shifts to perform an exact phase correction. On the 
other hand, the proposed Bayesian model of sections \ref{sec: modelspec}-\ref{sec: estothers} automatically learns phase 
corrections from the data.}  We simulate FID signals using the generative 
model of Equation~\eqref{eqn: generative}, using the theoretical
intensities from Table~\ref{tab:excitation-profile}.  Note that in real experiments, resonant frequencies will shift depending on the
composition of a mixture, especially in the presence of commonly used solvents such as CDCl$_3$.  Thus to make the simulations more 
representative of real experiments, and more challenging, we let the ground truth frequencies differ from the tabulated values by $\pm 3$ ppm, 
and initialise the estimation procedure of section \ref{sec: estothers} using the values in Table~\ref{tab:excitation-profile}.

As can be seen in Figure~\ref{fig: fidcompare}, 
a synthetic FID from Eq.~\eqref{eqn: generative} resembles an actual FID response to a 
2-butanone cyclohexane mixture.  In section \ref{sec: realexperiments} we then compare these models
on experimentally acquired FID signals.

%
\subsection{Stress Tests}

We start by stress testing the proposed model in response to varying signal to noise
ratios (SNR) (as defined in Eq.~\eqref{eqn: SNR}) in the FID signal.

\begin{figure}
\centering
\includegraphics[width=0.5\linewidth]{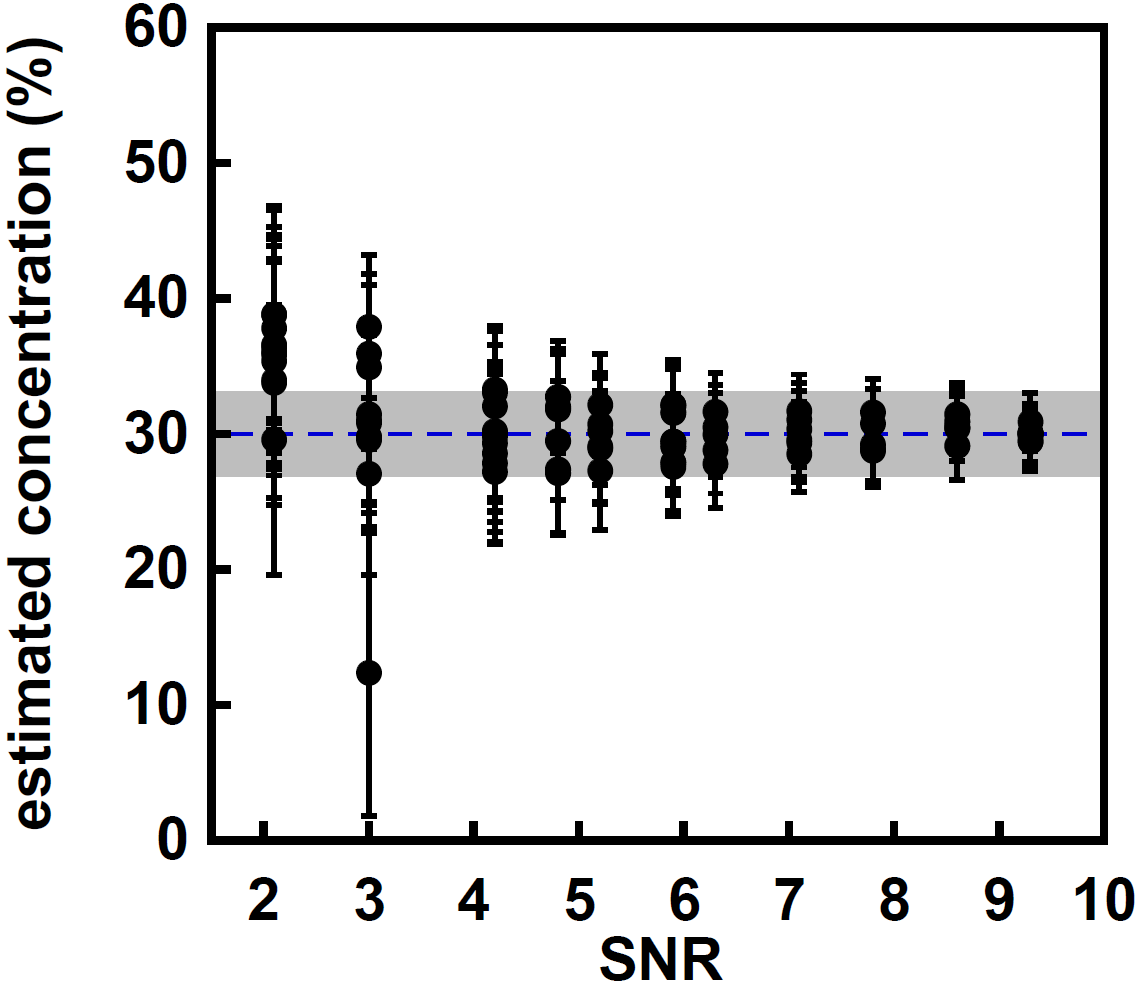}
\caption[]{
The relative concentration (and uncertainty) of 2-butanone in synthetic cyclohexane 2-butanone
mixtures at various SNR, using the proposed model.  The true concentration (30\%) is given by the dashed line, with the shaded
area representing 27\%--33\% 2-butanone.  The error bars respresent the $95\%$ credible region.
}
\label{fig:bayesian-ft-butanone30-snr}
\end{figure}

Taking a mixture of $30\%$ 2-butanone $70\%$ cyclohexane as an example,
estimates and uncertainties regarding the concentration of 2-butanone,
using the proposed model, are shown in Figure~\ref{fig:bayesian-ft-butanone30-snr}, 
with the true concentration of 2-butanone (30\%) given
by the dashed line.  The gray shade indicates a concentration between 27\%--33\%.  The error
bars indicate a $95\%$ credible interval (two sample standard deviations of $p(|A_i/A_j| | \bm{y}_1, \bm{y}_2, \psi, v)$ using $10000$ samples
and the proposed model of section \ref{sec: modelspec}).  

To test the robustness of the proposed model to noise, we selected a set of SNR values, and generated data 30 times for
each of these values. As shown in Figure~\ref{fig:bayesian-ft-butanone30-snr}, above a SNR of
$4$, the estimated concentrations are located within the
gray shade (27\%--33\% 2-butanone), and contain the true concentration (30\%) in 29/30 experiments. 
As expected, the estimated concentrations decrease in variance across different datasets as SNR increases,
and converge to the true generating values.  Figure~\ref{fig:bayesian-ft-butanone30-snr} also shows that below an SNR of
around $4$, the proposed model tends to overestimate the concentration of 2-butanone.

We now analyse the effect of increased noise on the marginal likelihood of the
data, while keeping all other parameters fixed.
Since the four resonant frequencies of 2-butanone behave similarly,
we analyse the frequency near $206.29$ ppm.

\begin{figure}
\centering
\subfigure[]{%
	\label{fig:3070-0-a}%
	\includegraphics[width=0.45\linewidth]{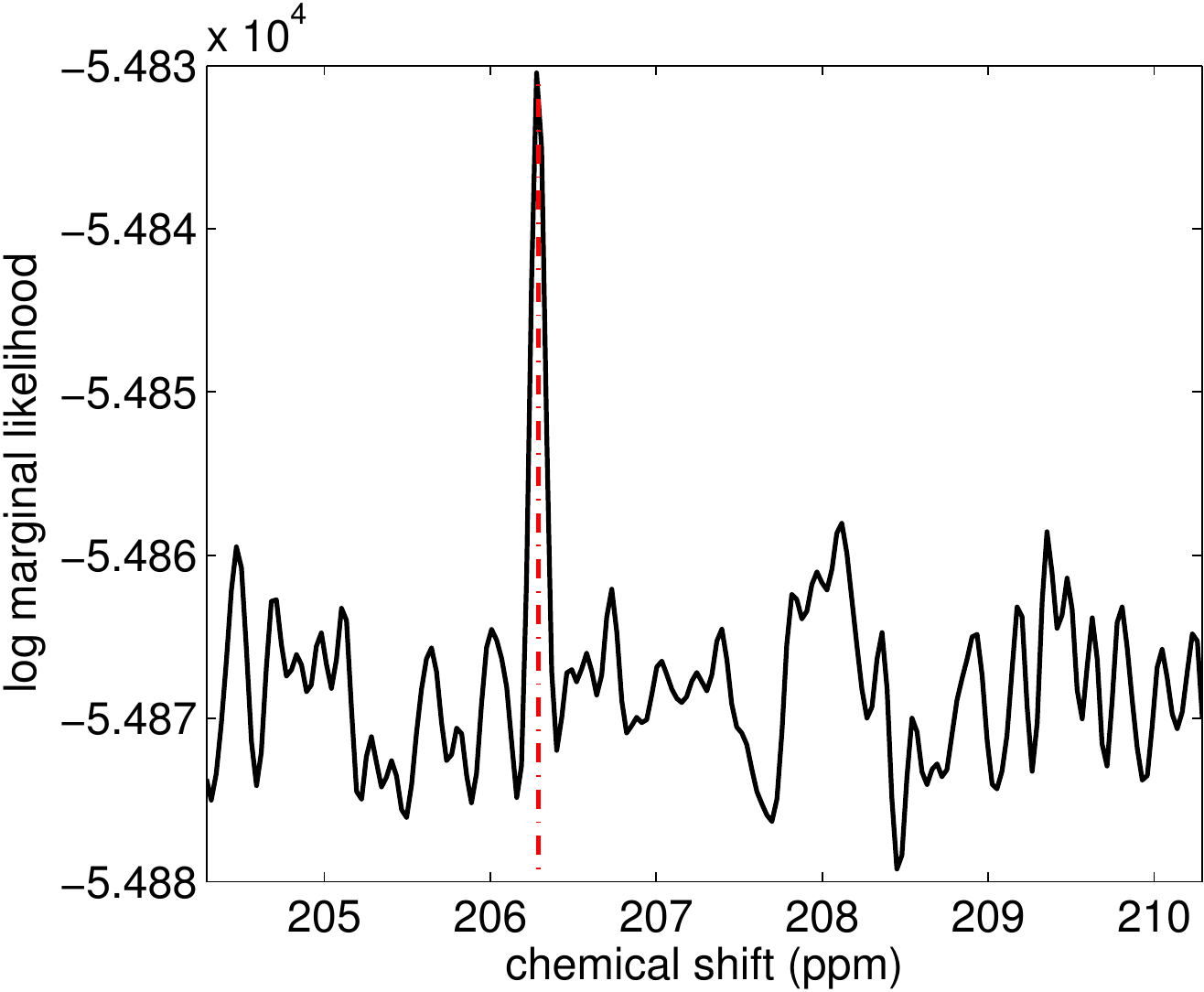}%
}%
\hspace{0.5cm}%
\subfigure[]{%
	\label{fig:3070-0-b}%
	\includegraphics[width=0.45\linewidth]{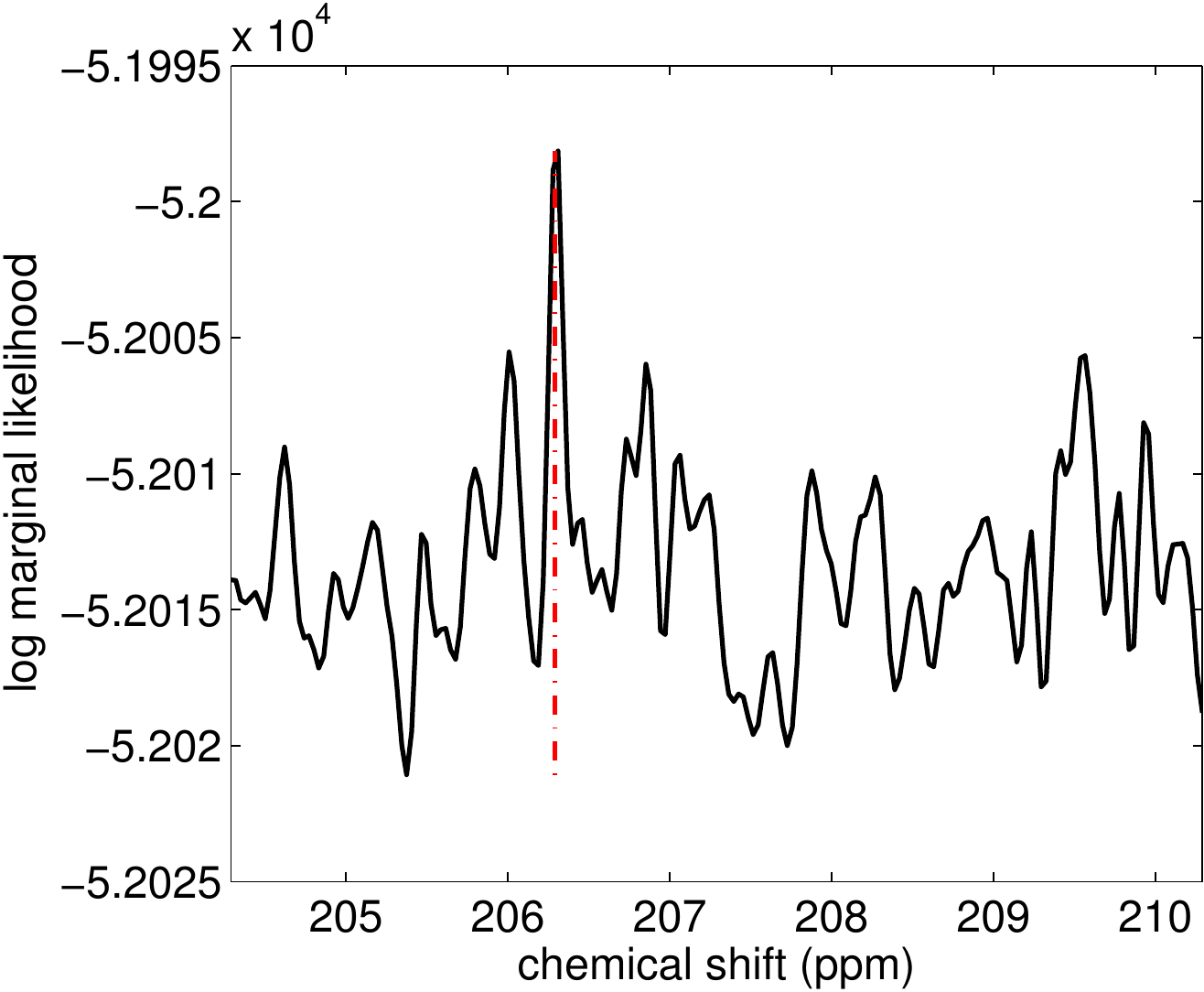}%
}%
\\%
\subfigure[]{%
	\label{fig:3070-0-c}%
	\includegraphics[width=0.45\linewidth]{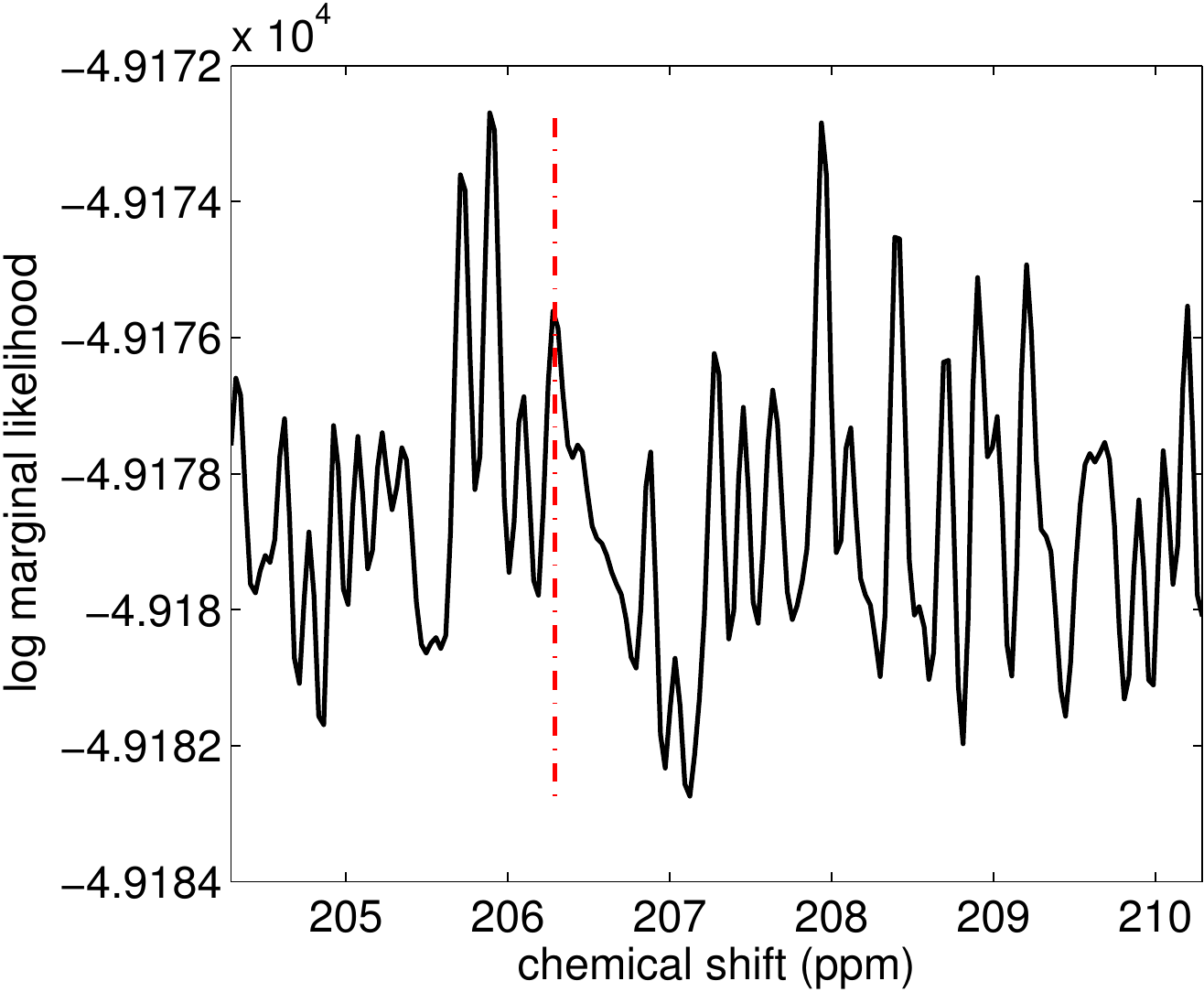}%
}%
\hspace{0.5cm}%
\subfigure[]{%
	\label{fig:3070-0-d}%
	\includegraphics[width=0.45\linewidth]{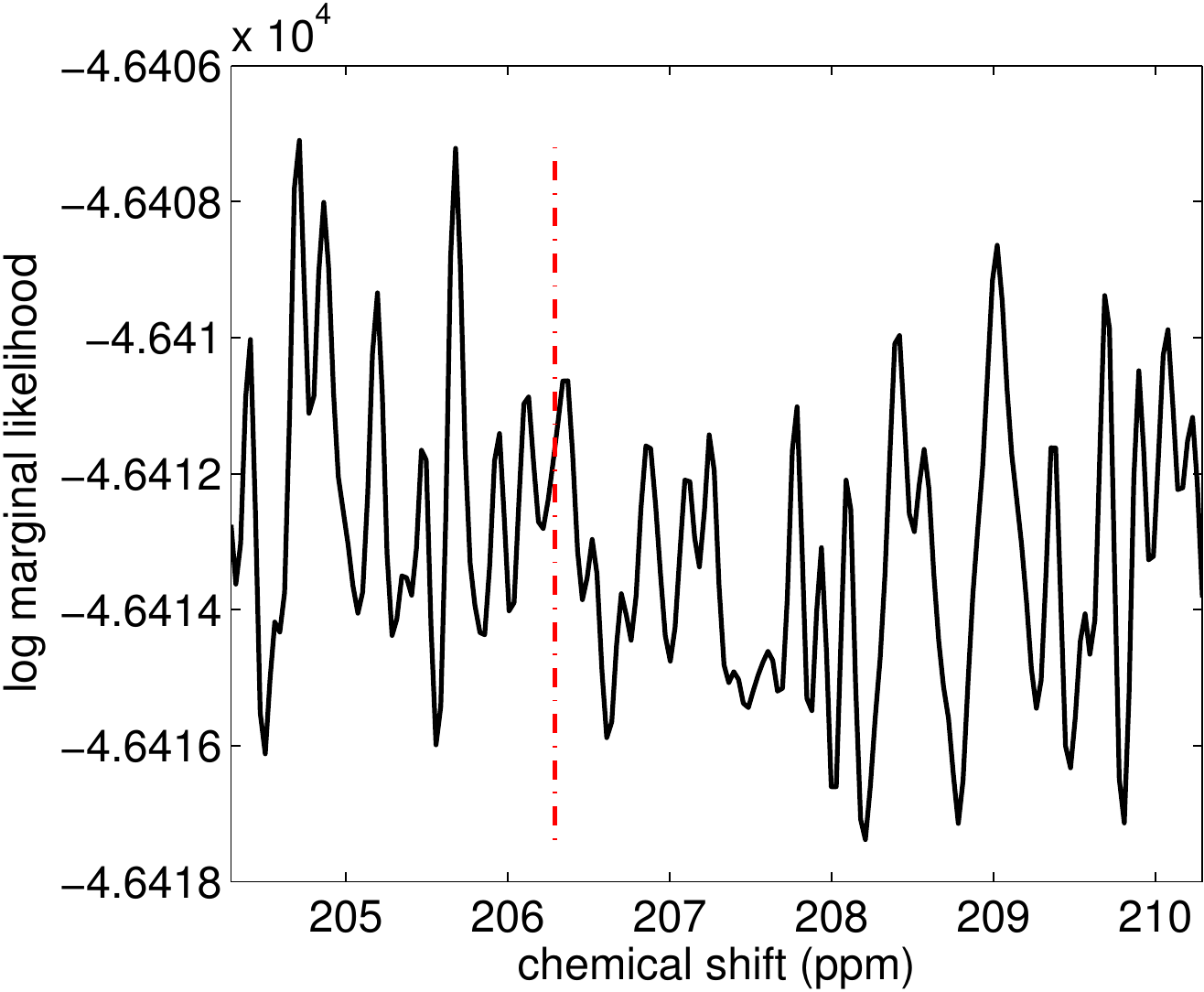}%
}%
\caption{
Log marginal likelihood plots as a function of a resonant frequency for butanone, with
the true frequency ($206.29$ ppm) indicated by the dashed line.    
The corresponding SNR for
each panel are
\subref{fig:3070-0-a} $5.9$,
\subref{fig:3070-0-b} $4.2$,
\subref{fig:3070-0-c} $3.0$, and
\subref{fig:3070-0-d} $2.1$.
For high SNR values,~\subref{fig:3070-0-a} and~\subref{fig:3070-0-b}, the peak
at the true frequency is clearly identifiable.
}
\label{fig:bayesian-noise-robustness}
\end{figure}

Figure~\ref{fig:bayesian-noise-robustness} displays the log marginal likelihood
as a function of the frequency. The SNR gradually decreases
from~\subref{fig:3070-0-a} to~\subref{fig:3070-0-d}, with $5.9$ for
\subref{fig:3070-0-a}, $4.2$ for~\subref{fig:3070-0-b},
$3.0$ for~\subref{fig:3070-0-c}, and $2.1$ for~\subref{fig:3070-0-d}.
As shown in Figure~\ref{fig:bayesian-noise-robustness}, the true
frequency (indicated by the dashed line) becomes less differentiated as the SNR
decreases, and the likelihood surface becomes increasingly multimodal, with no
global optimum near the true generating frequencies.  The precision of estimating the 
frequencies as well as the final concentrations at selected SNR were analysed by 30 repetitions 
and are given in Table~\ref{tab:bayesian-noise-robustness}.

In Figure~\ref{fig:3070-0-a} the true frequency is well-resolved, and indeed the proposed approach 
estimates the frequencies accurately, as given in 
Table~\ref{tab:bayesian-noise-robustness}. As the SNR decreases to $4.2$ (Figure~\ref{fig:3070-0-b}),
the marginal likelihood profile becomes less peaked at the true frequency.  At such an SNR,
the credible sets for the proposed method become relatively large, as shown in 
Table~\ref{tab:bayesian-noise-robustness}, and the method is more susceptible to the
local optima shown in the likelihood surface of Figure \ref{fig:marginal-likelihood-vis}.  At this SNR level, the
issue of local optima is largely alleviated with multiple SIMPSA restarts.

\begin{table}
\centering
\caption{Bayesian estimates of the 2-butanone ($30\%$) and cyclohexane
($70\%$) mixture.
}
\label{tab:bayesian-noise-robustness}
\begin{tabular}{c>{\centering\arraybackslash}m{3cm}>{\centering\arraybackslash}m{2cm}>{\centering\arraybackslash}m{3cm}>{\centering\arraybackslash}m{3cm}}
\hline
SNR & Error in frequency estimation (ppm) (mean $\pm 1$ std. dev.)
	& Bayesian $95\%$ credible interval
	& Empirical Coverage with $68\%$ and $(95\%)$ credible intervals
	& Percentage of reconstructions within 27\%-33\% 2-butanone
\\
\hline
$9.3$ & $-0.0039 \pm 0.0096$ & $1.09\%$ & $91\%$~$(100\%)$ & $100\%$\\
$5.9$ & $-0.0072 \pm 0.0043$ & $3.44\%$ & $50\%$~$(100\%)$ & $100\%$\\
$4.2$ & $-0.0067 \pm 0.0187$ & $4.92\%$ & $82\%$~$(100\%)$ & $82\%$\\
$3.0$ & $0.3383 \pm 0.7159$ & $6.80\%$ & $64\%$~$(73\%)$ & $60\%$\\
$2.1$ & $0.5608 \pm 1.6385$ & $8.34\%$ & $27\%$~$(82\%)$ & $9\%$\\
\hline
\end{tabular}
\end{table}

Figures~\ref{fig:3070-0-c} and \ref{fig:3070-0-d} show the marginal likelihood
at SNR values of $3.0$ and $2.1$ respectively.  
As the SNR increases, predictions with the proposed Bayesian model becomes increasingly accurate,
converging to the truth, as shown in the second and last
column in Table~\ref{tab:bayesian-noise-robustness}.  When the SNR is above
$4.2$, the proposed Bayesian method's $95\%$
posterior credible interval always contains the true concentration.

%

\subsection{Simulated Comparison to Conventional Spectroscopy: High Concentrations}
\label{subsec: simbvfhc}
We now examine the behaviour of both models in predicting high concentrations of 
a given chemical.

\begin{figure}
\centering
\includegraphics[width=0.5\linewidth]{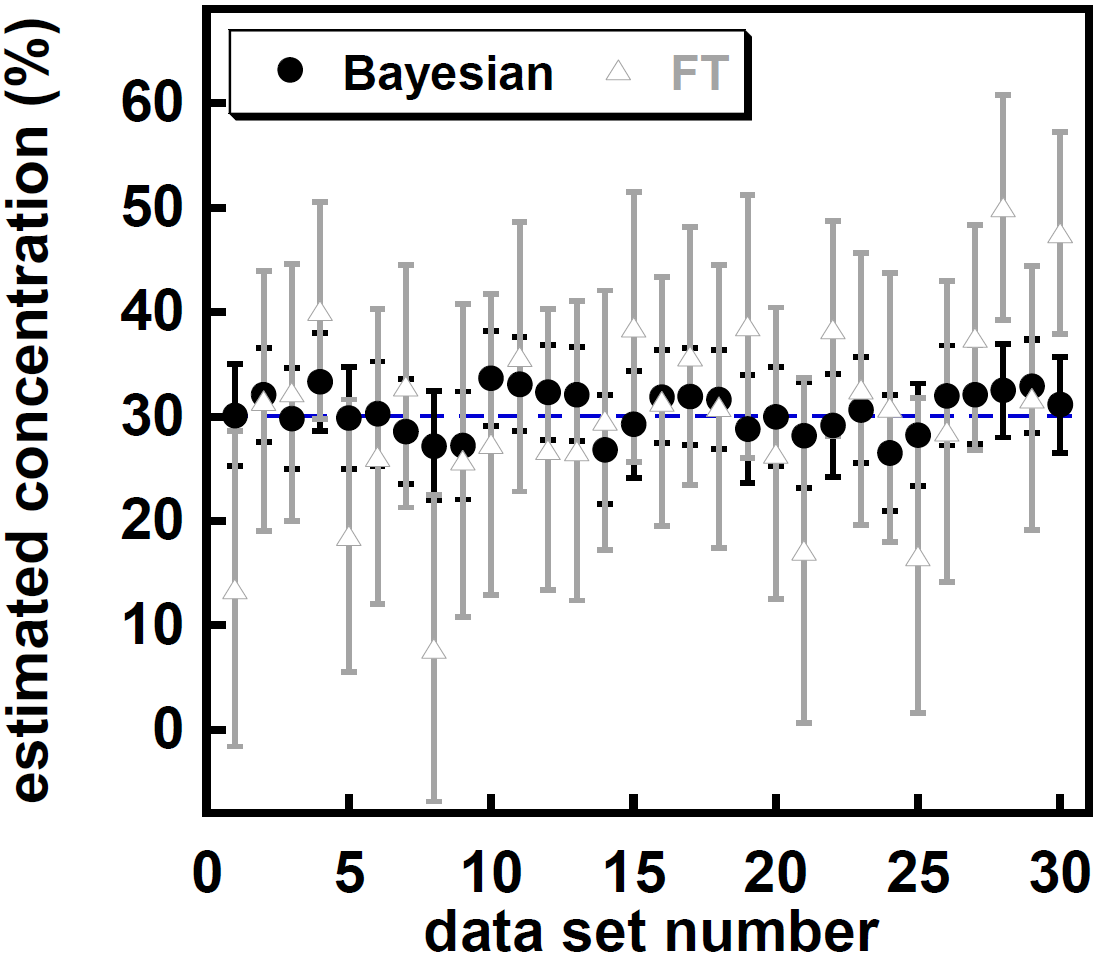}
\caption[]{Comparison of the Bayesian and FT performances on the same FID
datasets.  30 repetitions are generated for 30\% 2-butanone at an SNR of
$4.2$.  The dashed line indicates the true
concentration and the error bars correspond to $95\%$ credible regions.
}
\label{fig:bayesian-ft-butanone30}
\end{figure}

Results on a sample of $30\%$ 2-butanone at an
SNR of $4.2$ are shown in Figure~\ref{fig:bayesian-ft-butanone30}, with the
true concentration indicated by the dashed line and the $95\%$ credible region given
by the error bar. To confirm
reproducibility, datasets were generated 30 times with the same concentration
and SNR.
Figure~\ref{fig:bayesian-ft-butanone30} shows that the proposed approach
consistently makes accurate predictions that always include the true value in
the 95\% credible interval. Its estimates are always within a $30\%\pm3.5\%$
bound.

Conversely, predictions using the conventional FT approach vary significantly 
in a $\pm20\%$ concentration interval, and even with broader error bars than 
the proposed approach FT estimates are typically unable to cover the true 
concentration.  The FT approach is particularly sensitive to noise, which will 
become even more apparent when we study low concentrations.

\begin{figure}
\centering
\subfigure[]{%
	\label{fig:1090-a}%
	\includegraphics[width=0.45\linewidth]{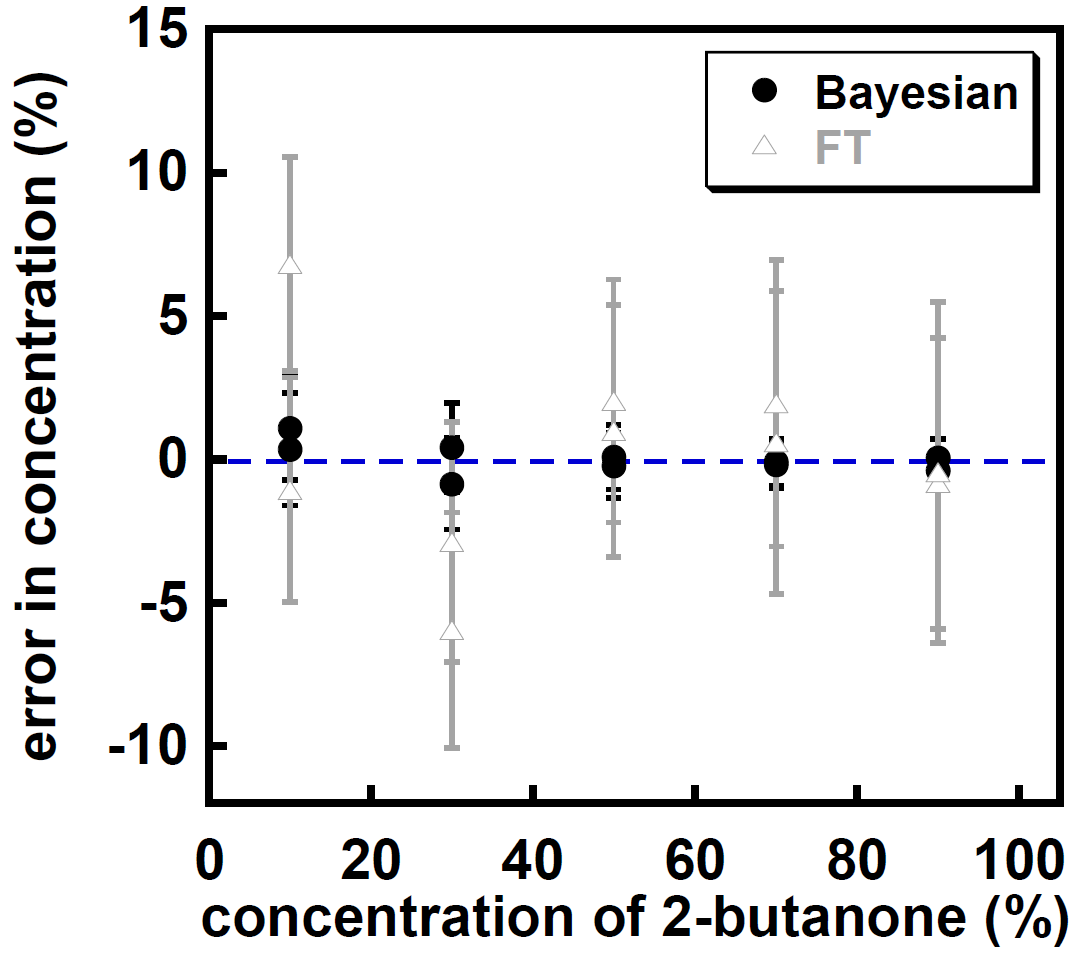}%
}%
\hspace{0.5cm}%
\subfigure[]{%
	\label{fig:1090-b}%
	\includegraphics[width=0.45\linewidth]{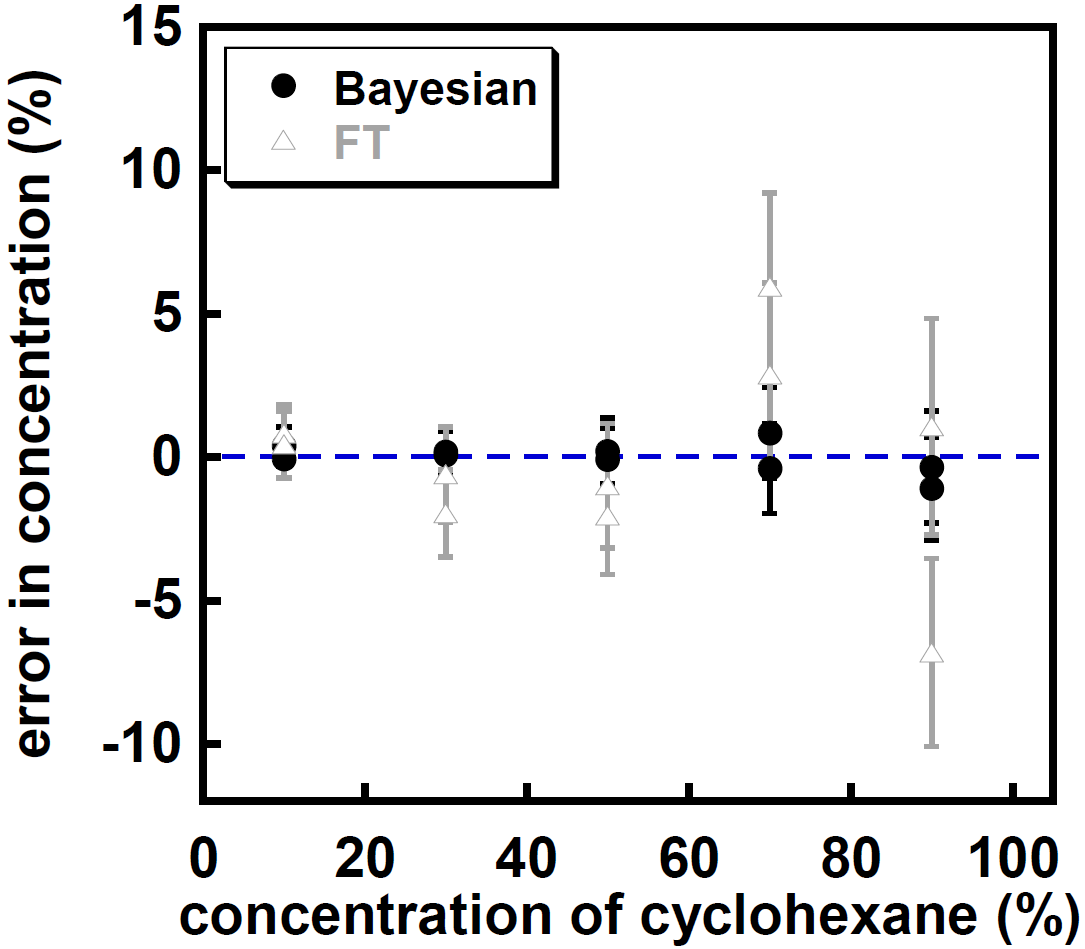}%
}%
\caption{
The proposed Bayesian and conventional FT estimations for varying relative concentrations of 
2-butanone and cyclohexane in simulation.  The
noise levels for all the data are the same and are chosen to allow the SNR to be as low as possible
but still retain acceptable performance.  In particular, the SNR is $4.5$ for the mixture containing
$10\%$ 2-butanone.  The vertical axis shows absolute errors.
The proposed Bayesian approach is more accurate and robust than the conventional FT approach in terms
of both expected mean and uncertainty.
}
\label{fig:bayesian-ft-1090}
\end{figure}

We perform a similar analysis for 2-butanone concentrations ranging from $10\%$ to
$90\%$. For clarity, representative results from only two datasets at
each concentration were plotted in Figure~\ref{fig:bayesian-ft-1090},
with the same noise level.\footnote{When we refer to \textit{absolute error}, we mean the absolute distance between the prediction and the true concentration percentage, and by \textit{relative error} we mean the ratio of the absolute error to the true concentration percentage.}

The Bayesian uncertainty intervals generally
decrease as the concentration of 2-butanone rises from $10\%$ to $90\%$.  Notice in 
Table \ref{tab:excitation-profile} that cyclohexane has a single intensity near $6$,
whereas 2-butanone has 4 intensities near $1$.  Due to the lower intensities of 
2-butanone, relative to cyclohexane, it is more difficult to 
differentiate the signal from $2$-butanone from the noise, and so predictions of
both cyclohexane and butanone concentrations
ought to be more accurate and certain as the relative concentration of $2$-butanone
rises.  Similarly, uncertainty
in the FT predictions of cyclohexane increase with relative increases in cyclohexane 
concentrations (decreases in 2-butanone).

In the cases presented, the proposed Bayesian approach consistently produces more accurate point predictions
and smaller and more accurate uncertainty intervals than a conventional Fourier transform approach.

\subsection{Simulated Comparison to Conventional Spectroscopy: Low
Concentrations}

We now investigate the performance of the proposed Bayesian approach compared to the
convential FT approach on low concentrations ($<10\%$). Low concentrations are
of particular interest, as spectroscopers often wish to determine the presence
or absence of a chemical in a system, or whether a particular chemical falls 
under a given concentration threshold.  As we will see, the lower the 
concentration of a chemical, the harder the concentration is to accurately
estimate.

\begin{figure}
\centering
\subfigure[]{%
	\label{fig:0595-a}%
	\includegraphics[width=0.45\linewidth]{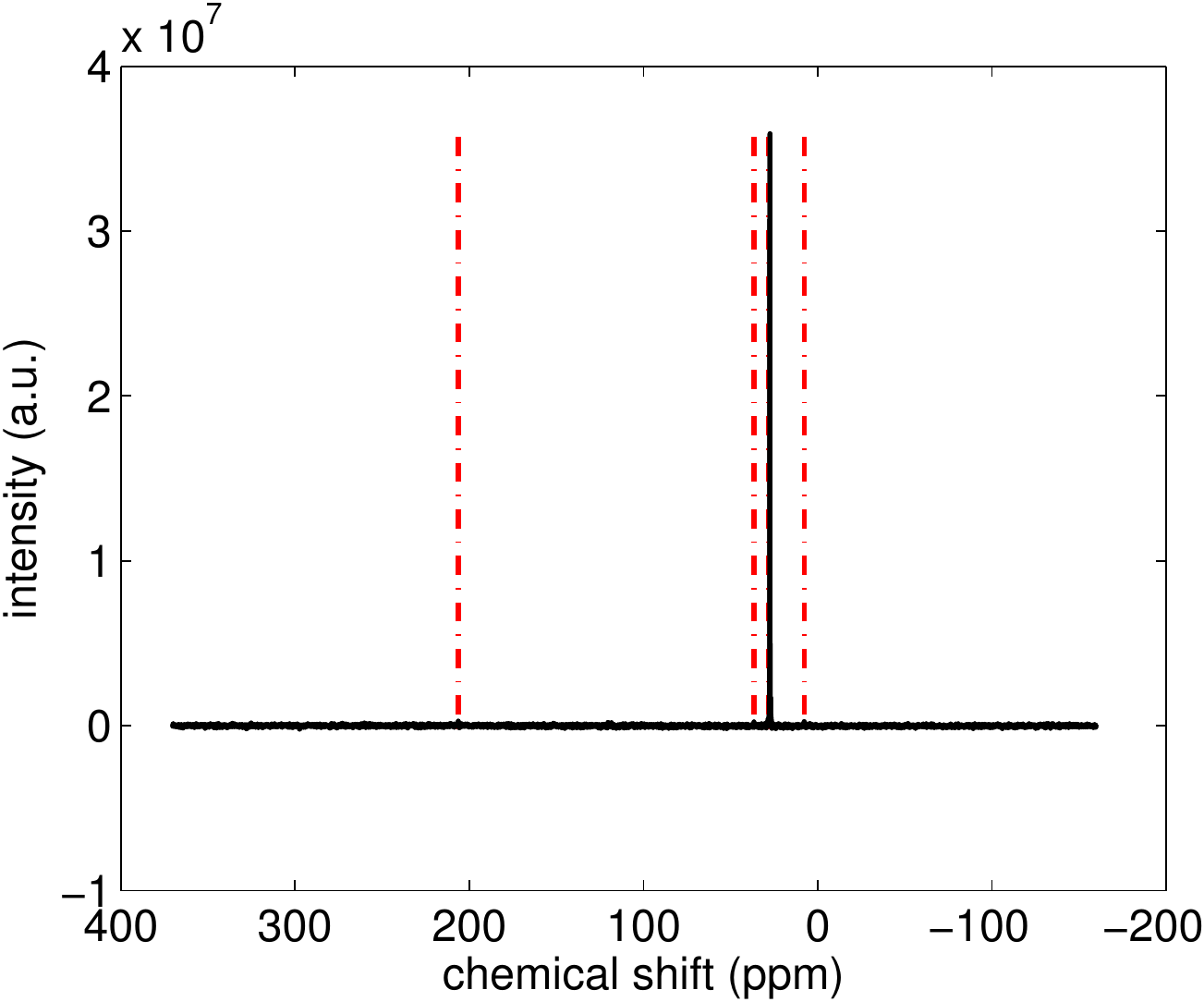}%
}%
\hspace{0.5cm}%
\subfigure[]{%
	\label{fig:0595-b}%
	\includegraphics[width=0.45\linewidth]{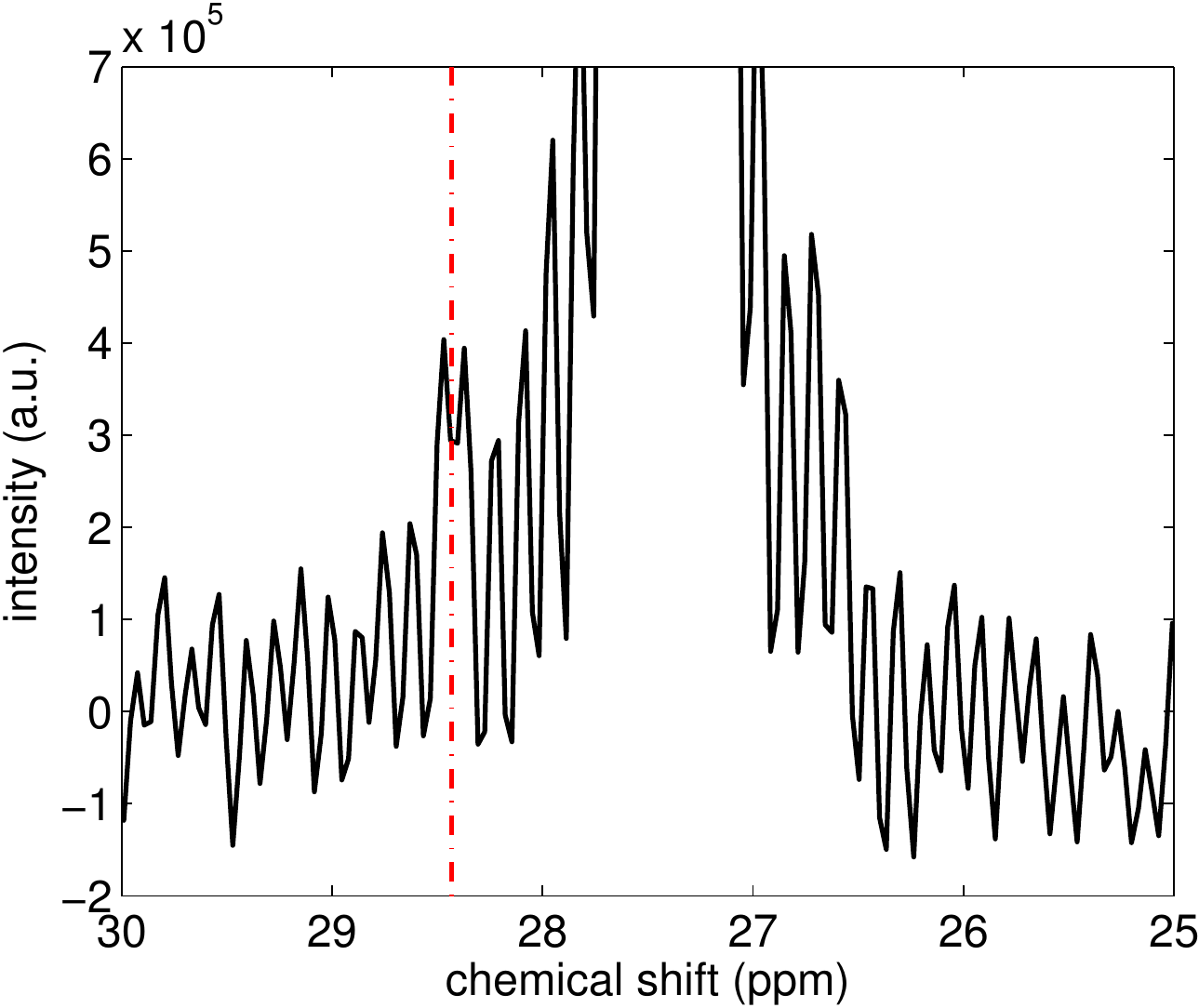}%
}%
\\%
\subfigure[]{%
	\label{fig:0595-c}%
	\includegraphics[width=0.45\linewidth]{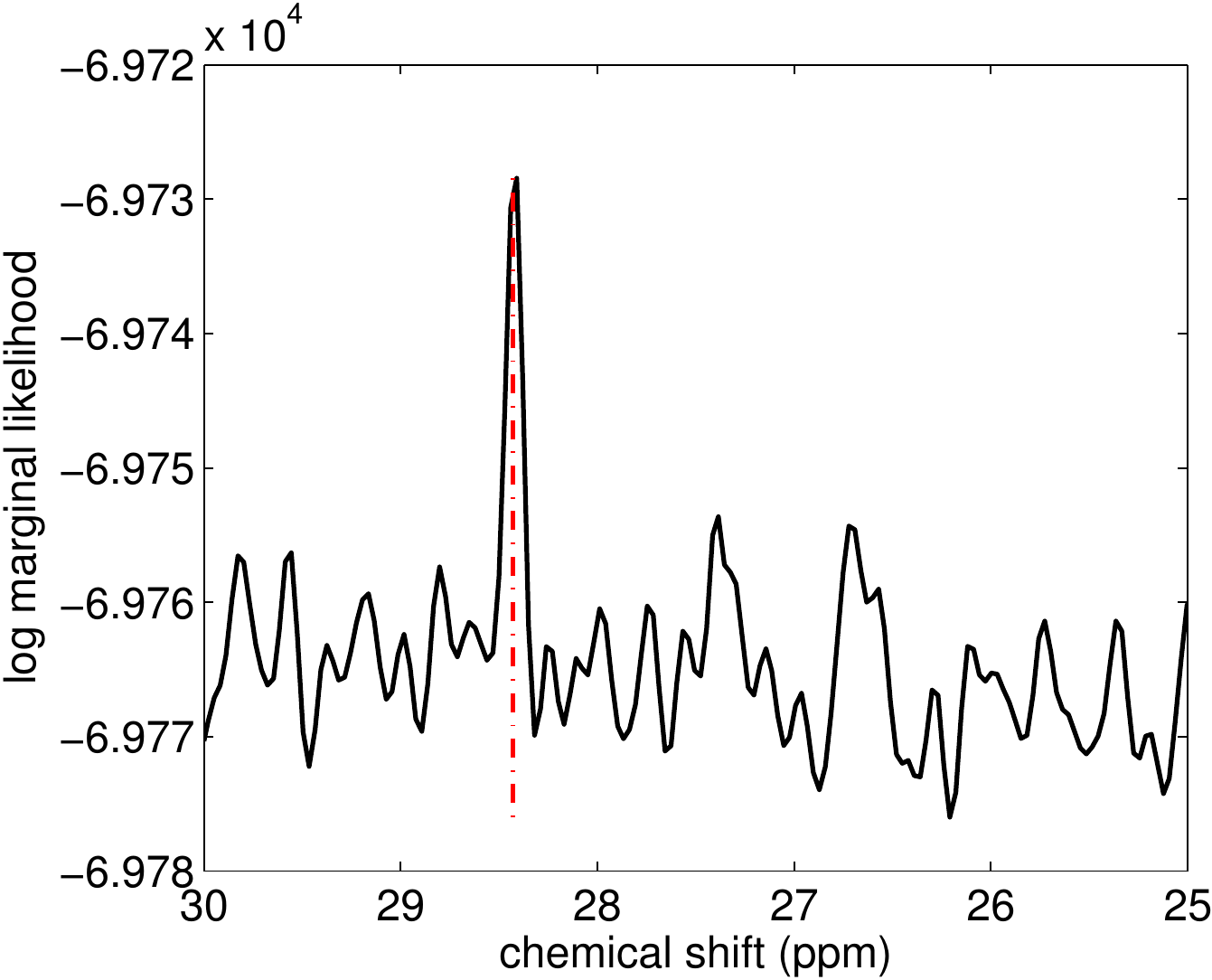}%
}%
\hspace{0.5cm}%
\subfigure[]{%
	\label{fig:0595-d}%
	\includegraphics[width=0.45\linewidth]{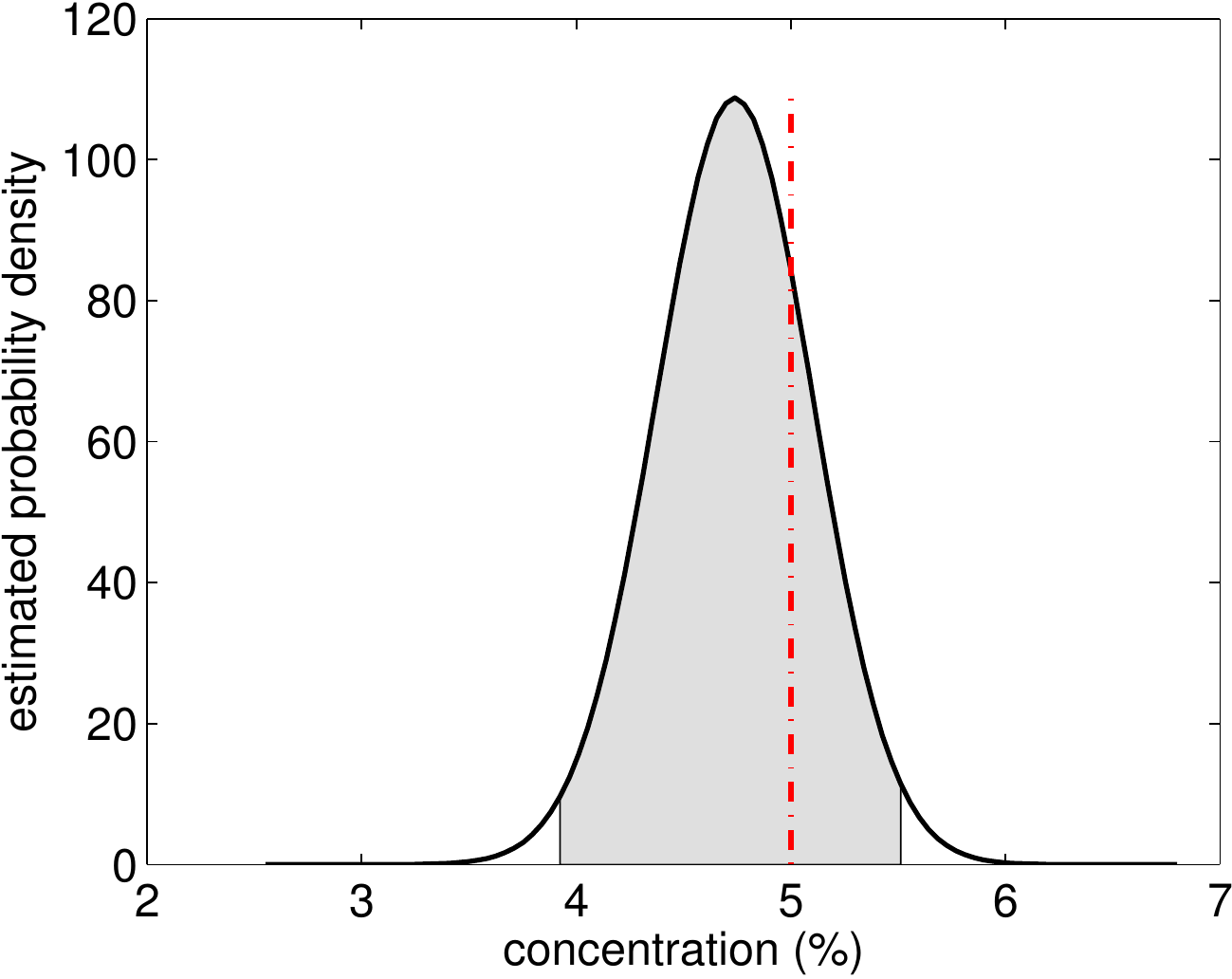}%
}%
\caption{
Comparison between the proposed Bayesian model and conventional Fourier transform spectroscopy on the same synthetic FID
dataset.  The ground truth frequencies 
are $206.29, 36.57, 28.43, 7.77$ ppm for butanone and $27.3$ ppm for cyclohexane.
The data were generated assuming a mixture of $5$ percent 2-butanone
with an SNR of $5.9$ 
\subref{fig:0595-a} The whole FT spectrum.
\subref{fig:0595-b} Zoom in to part of the spectrum indicated by the red box
in~\subref{fig:0595-a}. The peak ($28.43$ ppm) from 2-butanone is on the
shoulder of the peak ($27.3$ ppm) from cyclohexane, and is prone to overestimation owing to 
the truncation artifacts arising from that peak.
\subref{fig:0595-c} Log marginal likelihood as a function of the $28.43$ ppm
resonance frequency. It is well distinguished from adjacent frequencies.
\subref{fig:0595-d} the concentration distribution of 2-butanone estimated
from the proposed Bayesian model. The dashed lines indicate the true values.  As shown
in \subref{fig:0595-c} truncation artifacts do not interfere with the performance of the
proposed Bayesian model.
}
\label{fig:bayesian-ft-detail}
\end{figure}

We begin with a simulation using $5\%$ 2-butanone with an SNR of $6$.
Figure~\ref{fig:0595-a} shows the whole FT spectrum for this dataset, with the
locations of expected peaks indicated by the dashed lines.  It is very
difficult to see the low intensity peaks from 2-butanone.
Figure~\ref{fig:0595-b} zooms in to part of the spectrum indicated by the red box
in~\ref{fig:0595-a}. As shown in Figure~\ref{fig:0595-b}, a peak from the dilute species, 2-butanone, overlaps with a 
high intensity peak from the concentrated species cyclohexane, which makes assignment of this peak difficult, 
causing the concentration of dilute chemicals to be overestimated.  
Moreover, the duration of the acquisition is limited owing to the radio frequency (rf) pulse used for proton
decoupling \citep{keeler2011understanding}.  This limited acquisition time causes truncation artifacts on 
the cyclohexane peak at $27.3$ ppm, making it more difficult to differentiate the butanone peak at $28.4$ ppm.

The Bayesian log marginal likelihood as a function of the same frequency from
2-butanone is displayed in Figure~\ref{fig:0595-c},
with the same frequency window as used for the FT
in~\ref{fig:0595-b}.

As shown in Figure~\ref{fig:0595-c}, using the proposed Bayesian method, true resonant frequencies of dilute species are well-resolved even in the presence of a nearby intense peak,
due to the principled models of decay and noise in the FID signal.  Such log marginal
likelihood shape reduces the difficulty of learning the right frequency in the
Bayesian procedure and, as shown in Figure~\ref{fig:0595-d}, the final
probability distribution is well resolved with a $2.4\%$ relative error
($0.12\%$ absolute error).

\begin{figure}
\centering
\includegraphics[width=0.5\linewidth]{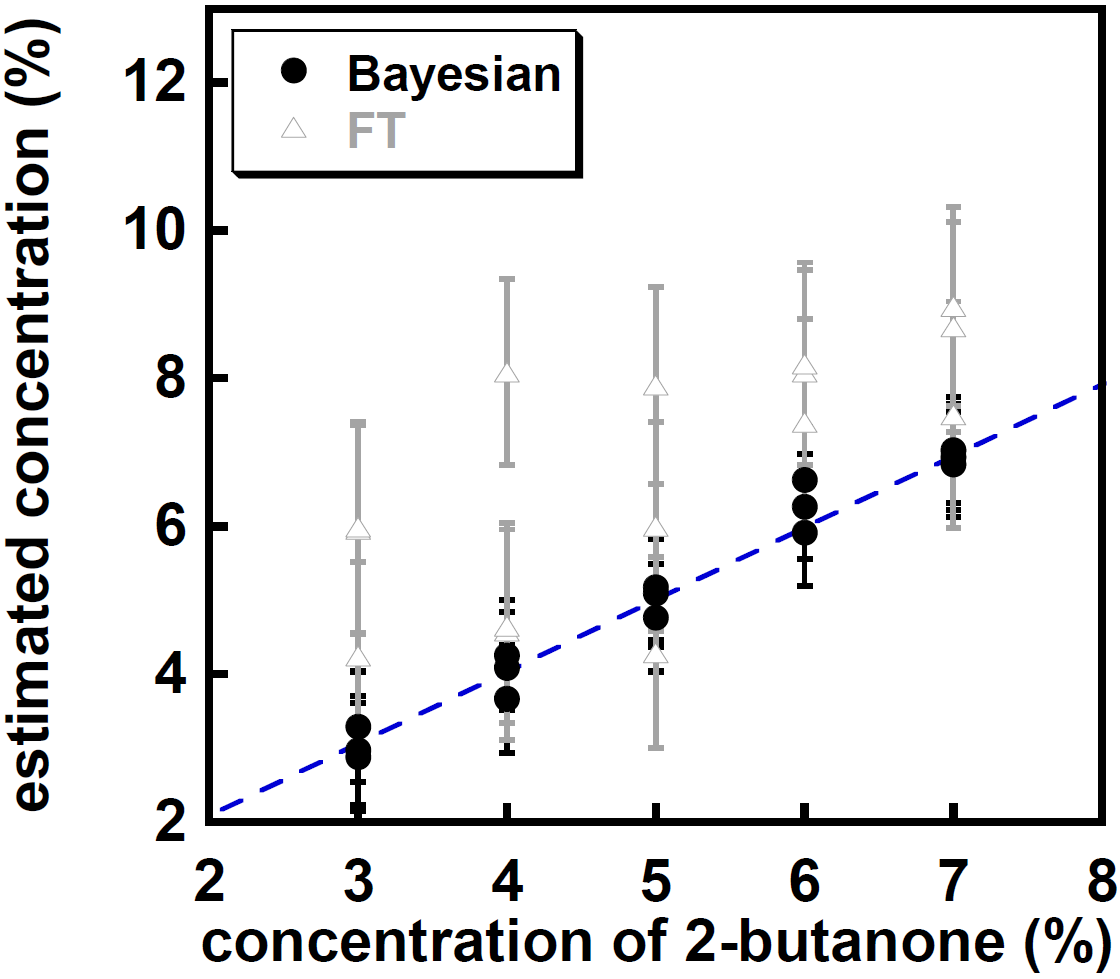}
\caption[]{
Simulated comparison of the Bayesian and FT approaches at low concentrations of 
2-butanone.
FID data were simulated with the a minimum noise level for acceptable performance
(an SNR of $4.4$, which is the smallest SNR at which the $3\%$ concentration can be
reasonably estimated).  The true
concentrations are given by the dashed line.
}
\label{fig:bayesian-ft-butanone-3to7}
\end{figure}

%
Simulated results
for $3\%$ to $7\%$ 2-butanone are shown in Figure~\ref{fig:bayesian-ft-butanone-3to7},
with the true concentration given by the dashed line. The FT predictions are
more variable than the Bayesian estimates, and systematically overestimate the
concentration in almost all cases.  Conversely, the Bayesian results have
no noticable bias and the true concentrations are always within the credible
region. The smaller error bar ($0.7\%$) -- which has been reliable in these
experiments -- also allows the Bayesian approach to reliably
distinguish between small concentration differences.

In conclusion, the analysis of the synthetic data at low concentrations
($<10\%$) demonstrates the strength of the Bayesian approach: it is more
accurate than the FT, particularly in peak overlapping situations, or when the
SNR or chemical concentration levels are particularly low.


\section{Experiments}
\label{sec: realexperiments}

We now compare the proposed Bayesian model with a conventional spectroscopy approach
on experimentally acquired FID signals corresponding to mixtures of cyclohexane and 2-butanone.

Cyclohexane ($\geq$ 99.5\%) and 2-butanone ($\geq$ 99.5\%) were purchased from
Sigma Aldrich. The mixtures were prepared by mixing the two chemicals with
various concentrations based on weight.  To confirm reproducibility of 
results, each mixture was made twice and both preparations were used in the experiments.
The error in sample preparations was calculated from the uncertainty in the electronic scale.

Conventional spectroscopy experiments were performed on a Bruker DMX300 spectrometer operating at a $^{13}$C
resonance frequency of 75.47 MHz with a 5mm microimaging probe. 
The reference frequency $\omega_0$ was calibrated using the standard reference tetramethylsilane (TMS).

The default digital filter on the spectrometer performs an unknown
transformation (for proprietary reasons) on the FID signal, making the 
generative model in Eq.~\eqref{eqn: generative} inapplicable.  
We therefore acquire the FID signal digitally, to minimise the noise in acquisition, and then use 
the TopSpin (Bruker) software on the spectrometer to convert the signal to a ``pseudo-analogue'' form.  This approach maintains the 
integrity of the generative model in Eq.~\eqref{eqn: generative}.  We found the performance of the 
conventional Fourier transform method is equivalent using the digital or pseudo-analogue form of the
FID signal.

\begin{figure}
\includegraphics[width=0.65\linewidth]{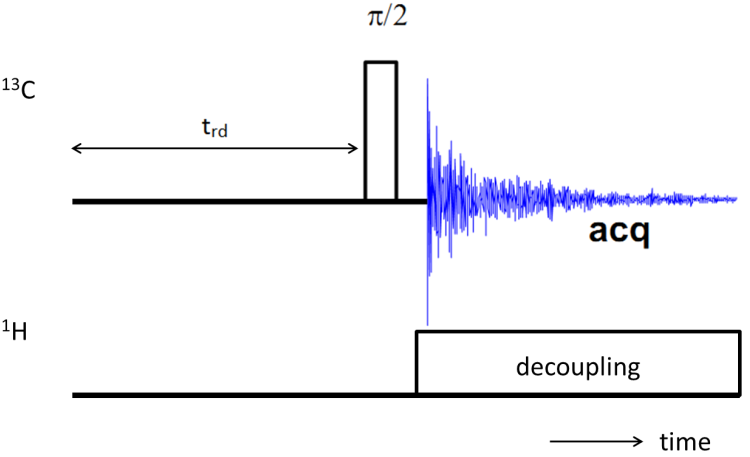}
\caption[]{
NMR on-pulse sequence.
}
\label{fig:fid-onpulse}
\end{figure}

Experimentally, the time domain FID data were acquired using a single
$90^{\circ}$ pulse on $^{13}$C channel with broadband $^1$H decoupling, as shown in
Figure~\ref{fig:fid-onpulse}.  Each FID data set consists of $4096$ complex
pairs with a $25 \mu s$ sampling interval. The recycle time (the time between
excitations of the system) was set as
$180s$ to allow sufficient relaxation recovery. The digitally filtered FID data
were converted to pseudo-analogue data format after acquisition using TopSpin 
software, which modified the
number of data points to $4029$.

Phase correction was manually performed on the spectrometer using both zero 
and first order phase corrections ($\theta$ and $\tau$ in Eq.~\eqref{eqn: generative})
to ensure all peaks are in the so-called ``absorption mode'' \citep{keeler2011understanding}.

Because the NMR spectrometer can only excite the spins within a certain resonance
frequency range, the extent to which each chemical group is excited differs
from each other slightly.  To account for this behaviour, we adjust the intensities
$B^{(k)}_i$ in the signal model of Eq.~\eqref{eqn: generative} as described in 
section \ref{sec: convspec} and Table \ref{tab:excitation-profile}.

%
\subsection{Experimental Comparison to Conventional Spectroscopy: High
Concentrations}

\begin{figure}
\centering
\subfigure[]{%
	\label{fig:exp-1090-butanone}%
	\includegraphics[width=0.45\linewidth]{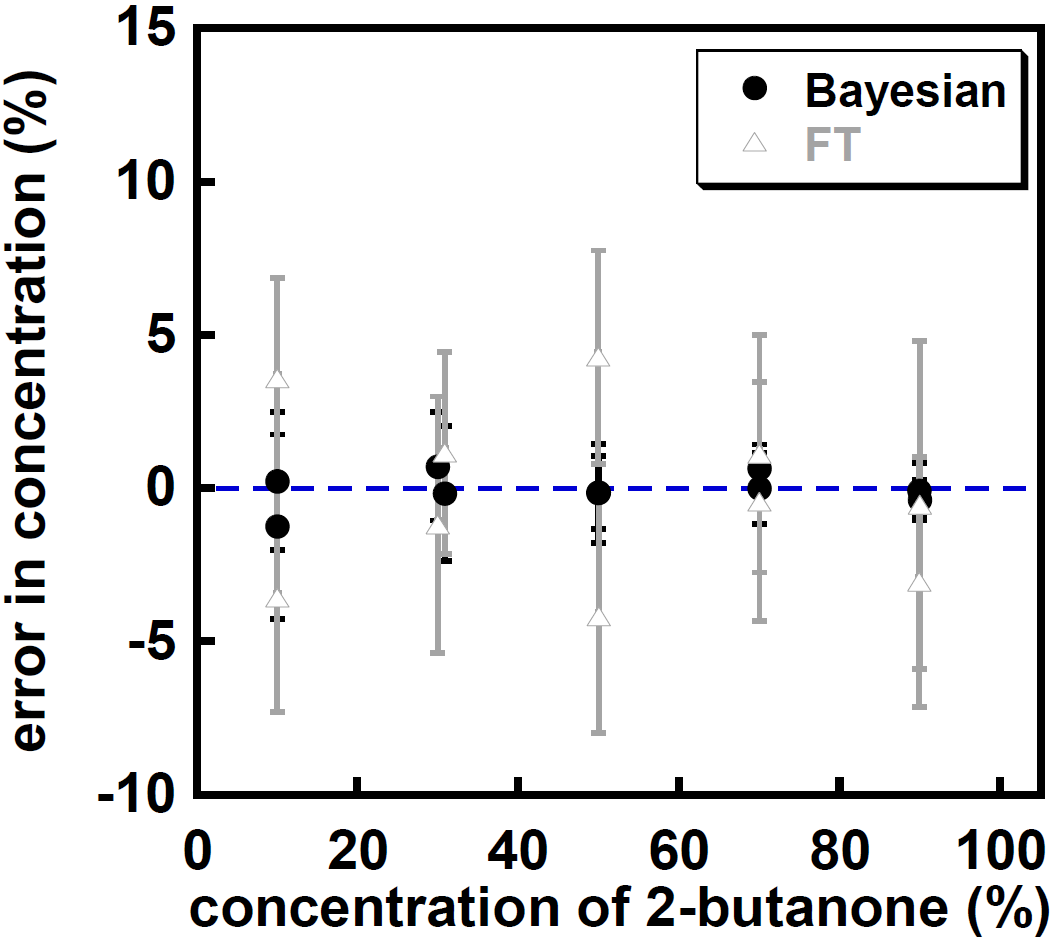}%
}%
\hspace{0.5cm}%
\subfigure[]{%
	\label{fig:exp-1090-cyclohexane}%
	\includegraphics[width=0.45\linewidth]{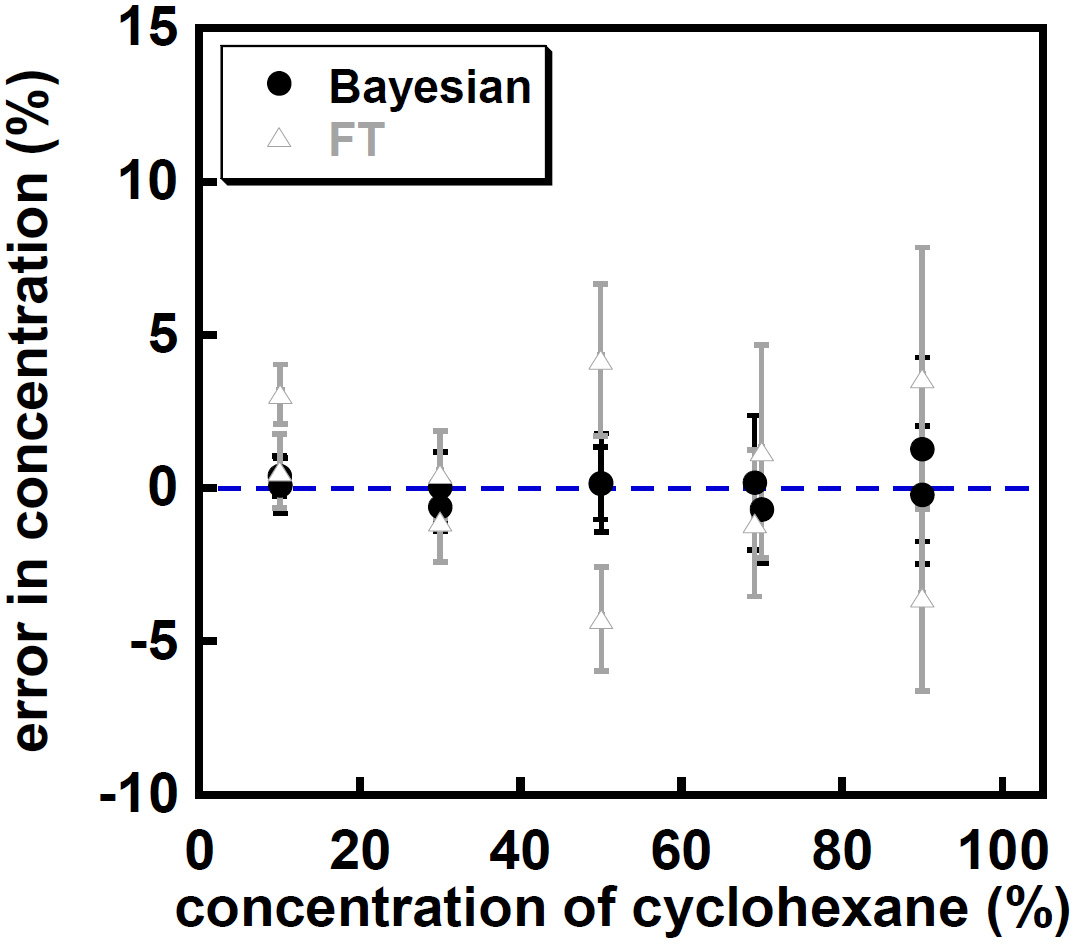}%
}%
\caption{
Experimental results from Bayesian and FT approaches.
The concentrations and SNR were set similar to the simulation cases shown in
Figure~\ref{fig:bayesian-ft-1090}.
To avoid bias in the ground truth, data were acquired on two sets of samples.
Error bars along the x axis were plotted but are too small to be seen.
}
\label{fig:exp-bayesian-ft-10-90}
\end{figure}

We first perform experimental measurements on mixtures of reasonably high
concentrations, from $10\%$ to $90\%$.  Similar to the simulations, the noise level
was set by fixing the number of scans in experiments, with an SNR for
the lowest concentration just above a threshold of 4.5, determined in section \ref{subsec: simbvfhc}.
To confirm the ground truth of the concentration, data were acquired from two
prepared samples at each concentration.  Estimation errors are shown in
Figure~\ref{fig:exp-bayesian-ft-10-90}.
Similar accuracy and general behaviour as in the simulations (Figure~\ref{fig:bayesian-ft-1090})
are observed.  Accordingly, the proposed Bayesian approach has far less error and more realistic
uncertainty estimates than the conventional Fourier transform approach.

%
\subsection{Experimental Comparison to Conventional Spectroscopy: Low
Concentrations}

\begin{figure}
\centering
\includegraphics[width=0.5\linewidth]{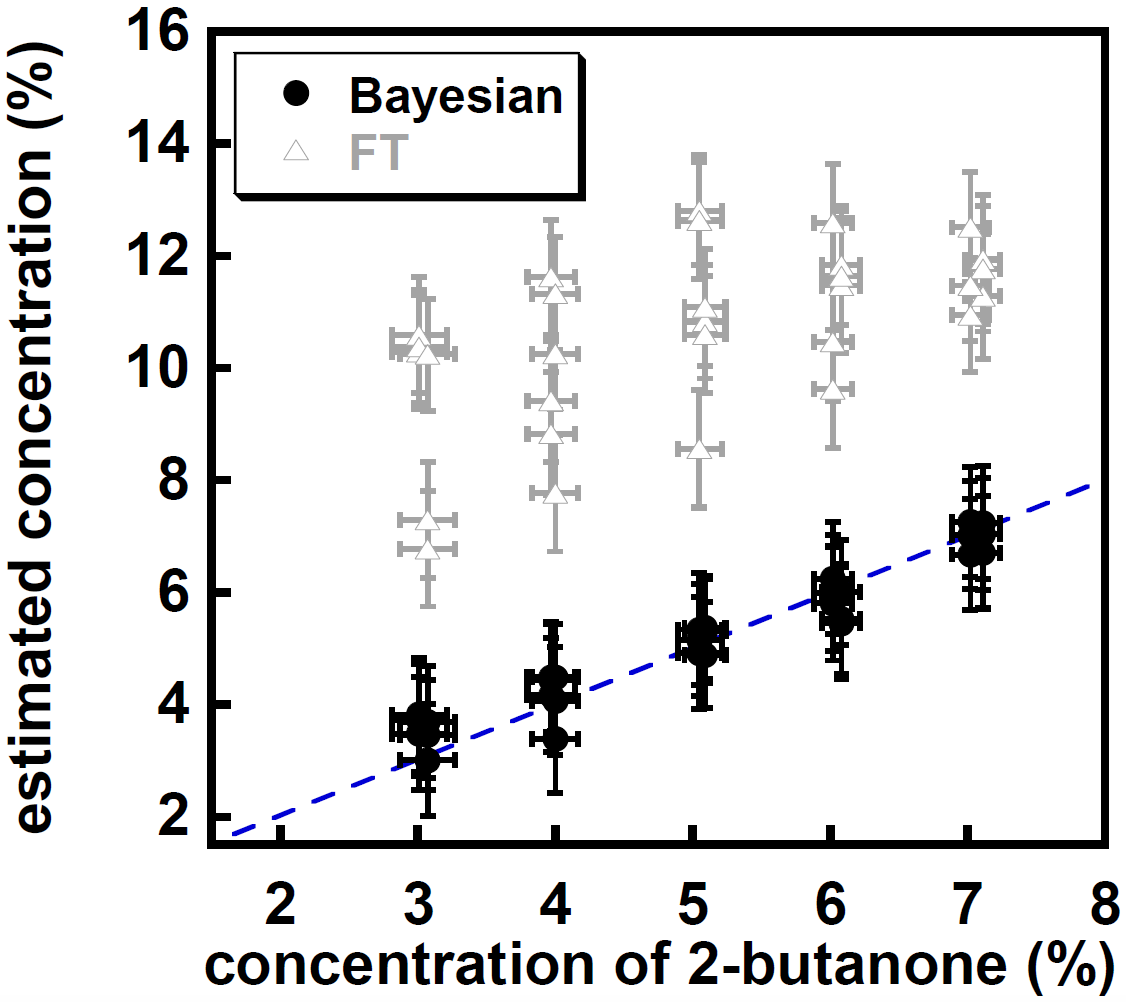}
\caption[]{
Experimental comparison between the Bayesian model and FT for inferring low
concentration levels.
The data were acquired at the same noise level that allows the lowest
concentration to be robustly estimated. Two samples were prepared at each
concentration with three independent datasets acquired from each sample.
}
\label{fig:exp-bayesian-ft-lowconc}
\end{figure}

We now investigate experimentally the ability of the proposed method and
conventional spectroscopy to estimate low concentration
levels of 2-butanone.
We prepare sets of samples with concentrations ranging from $3\%$ to $7\%$
in $1\%$ increments.  We create two sets per concentration level and perform
$3$ independent measurements on each sample.
The SNR for the $3\%$ cases is 4.4.
The Bayesian and FT
results are shown in Figure~\ref{fig:exp-bayesian-ft-lowconc},
with the expected concentration (ground truth determined during mixture preparation) indicated by the dashed line. The error bars
along the horizontal direction represent the error in sample preparation. The 
FT method consistently overestimates the concentration whereas there is no
systematic bias from the Bayesian results.  Moreover, the FT results from the
same sample can scatter over a $5\%$ range and their error bars are
unrealistically small.
In contrast, the Bayesian predictions are always within a $1\%$ concentration
bound and with more reliable uncertainty estimates.

\begin{figure}
\centering
\subfigure[]{%
	\label{fig:1161-a}%
	\includegraphics[width=0.31\linewidth]{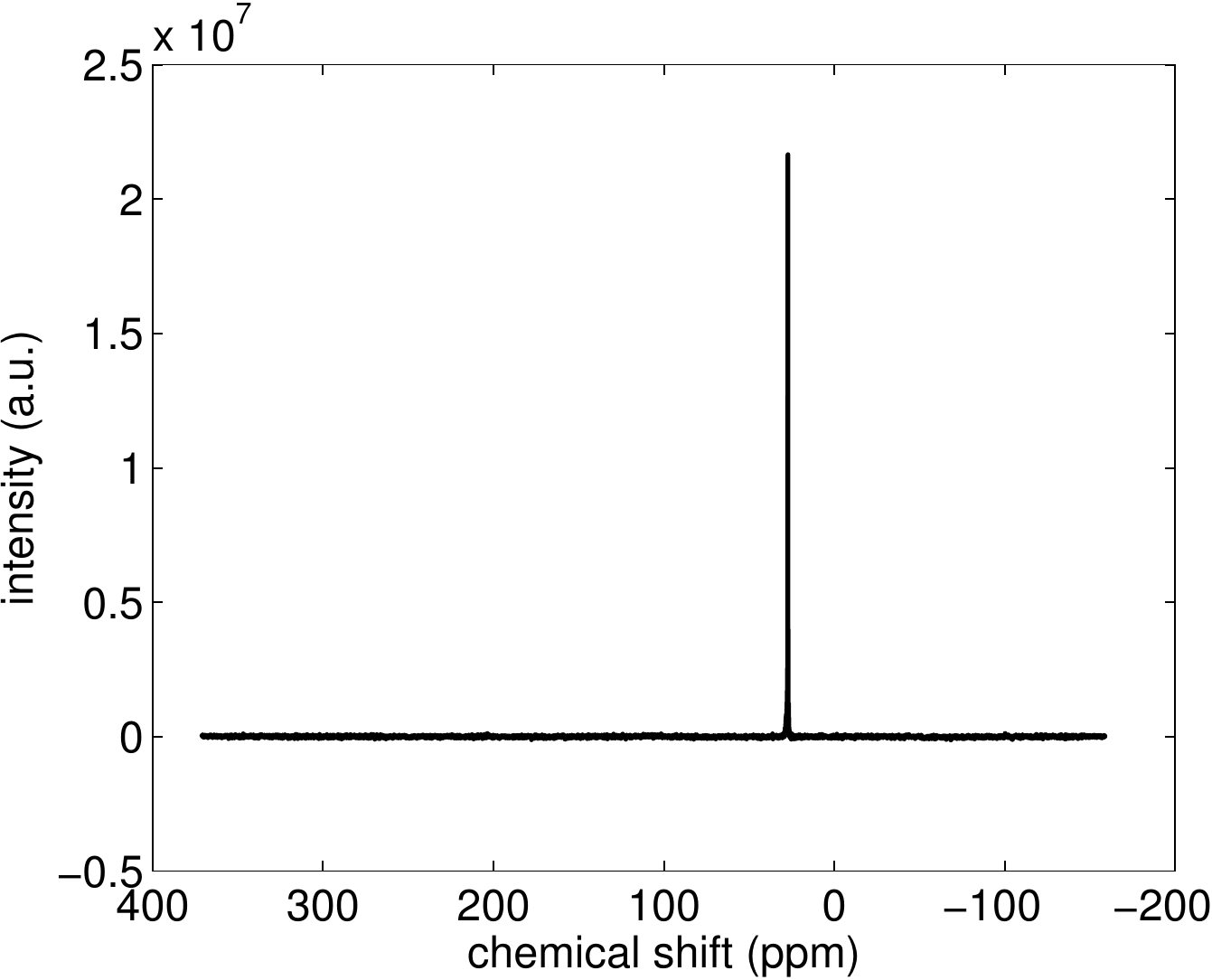}%
}%
\hspace{0.5cm}%
\subfigure[]{%
	\label{fig:1161-b}%
	\includegraphics[width=0.31\linewidth]{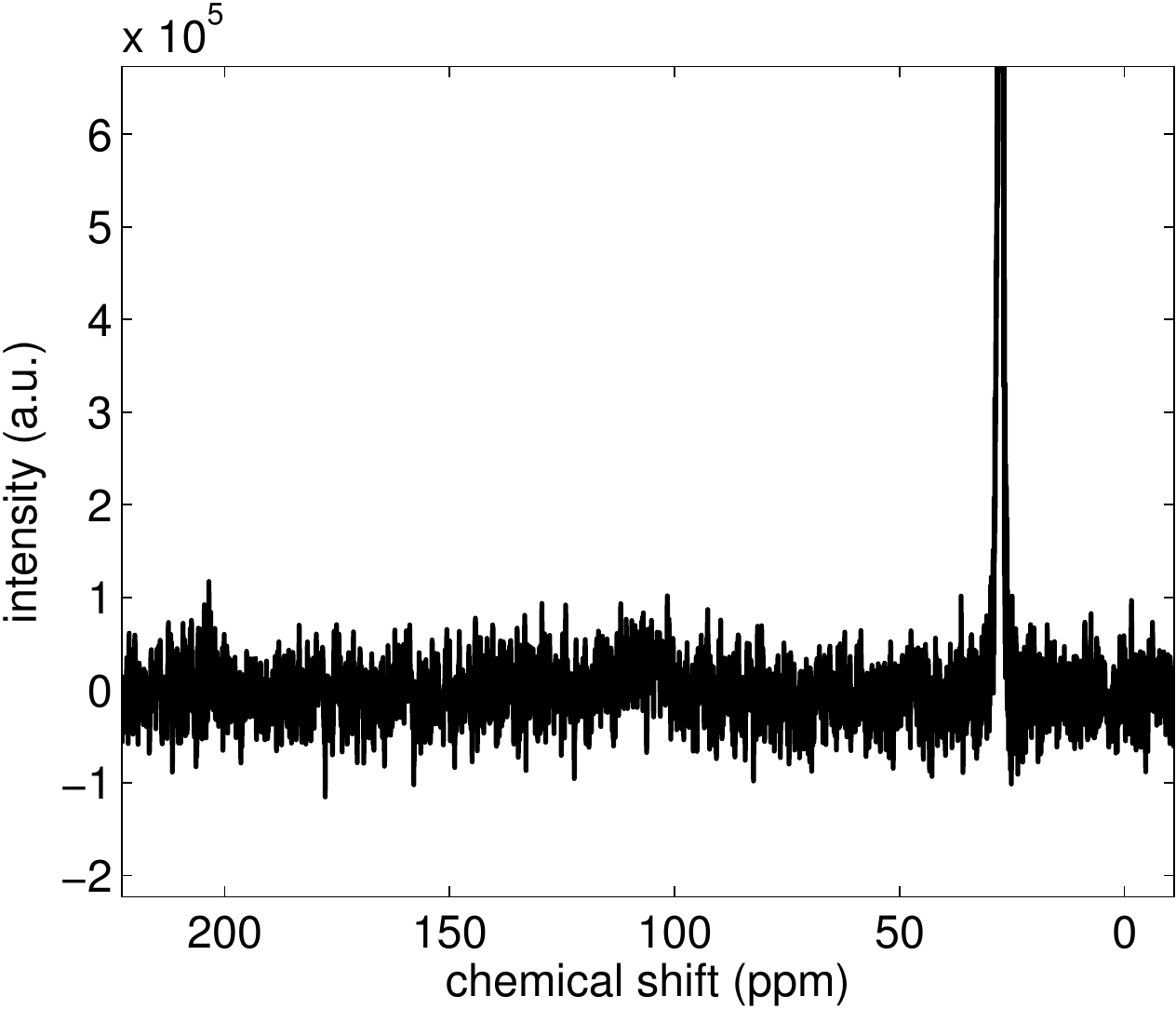}%
}%
\hspace{0.5cm}%
\subfigure[]{%
	\label{fig:1161-c}%
	\includegraphics[width=0.31\linewidth]{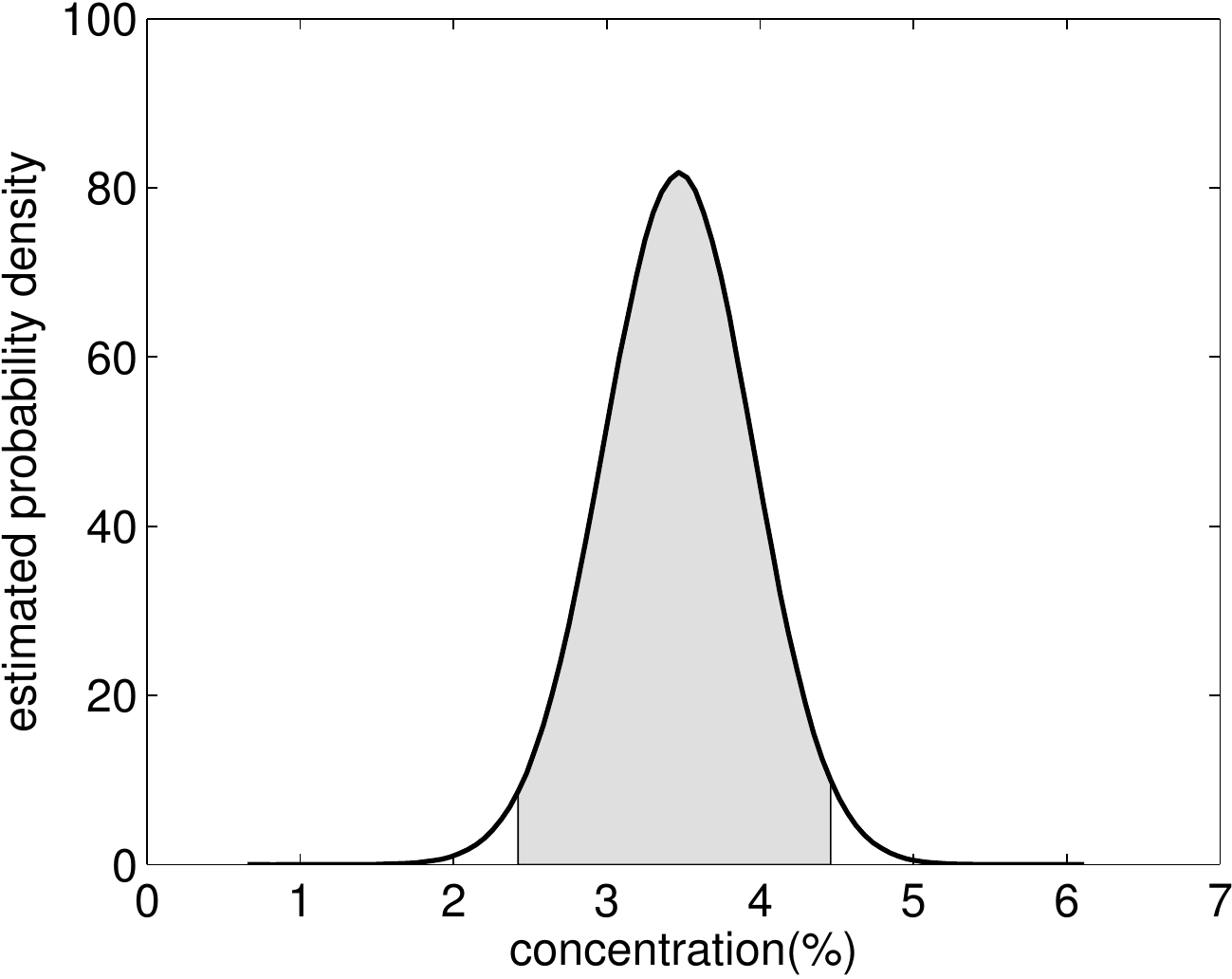}%
}%
\caption{
Experimental findings on a 
$3.07\%\pm 0.20\%$ 2-butanone mixture.
\subref{fig:1161-a} Conventional Fourier spectrum.
\subref{fig:1161-b} Zoom in to the low intensity 2-butanone peaks. The peaks
are almost at the same height as the noise level and cannot be assigned and
quantified accurately.
\subref{fig:1161-c} Concentration probability distribution of 2-butanone
calculated from the Bayesian model. The shaded area gives the $95\%$ credible
region.
}
\label{fig:exp-lowconc-detail}
\end{figure}

A particular set of data from the low concentration cases were taken to
provide a more visual description.  Figure~\ref{fig:exp-lowconc-detail}
shows the results of a dataset from a $3\%$ 2-butanone mixture.
The conventional FT spectrum, Figure~\ref{fig:1161-a}, does not show the peaks
from the $3\%$ 2-butanone. It is difficult to differentiate most of the peaks
in 2-butanone, as
shown in Figure~\ref{fig:1161-b}.  However, the proposed Bayesian approach can still
infer the true concentration with reasonable accuracy and uncertainty.

\section{Discussion}
\label{sec: discussion}

We have proposed an alternative method for nuclear magnetic resonance spectroscopy.  Unlike conventional Fourier transform spectroscopy,
the proposed approach explicitly models decay, noise, frequencies, intensities, and phase shifts, and can leverage prior information in 
a principled way using probability distributions.  We compared the proposed Bayesian approach with conventional Fourier transform 
spectroscopy in simulations and on experimentally acquired free induction decay signals, for quantifying relative chemical concentrations
in a mixture.  The proposed approach significantly outperformed conventional spectroscopy, with more accurate estimates and uncertainty 
intervals, particularly in low signal to noise ratio cases, and in cases where there were overlapping peaks in the Fourier transform of 
the free induction decay.  In summary, the proposed method -- while sensitive to frequency estimation -- can be used with little human
intervention, for reproducible and accurate estimates of chemical concentrations, and may be used in detecting the presence or absence
of chemicals where conventional Fourier transform spectroscopy cannot be used.  In general, the robustness of the proposed method to 
low signal to noise ratios, and overlapping spectral peaks, has widespread promise in analytic chemistry.

\section*{Acknowledgements}

The authors would like to thank NSERC, Microsoft Research Connections, and the EPSRC (Grants No. EP/F047991/1 and EP/K039318/1) for financial support.


\bibliography{paper}
\bibliographystyle{imsart-nameyear}

\end{document}